\documentclass{article}

\usepackage{arxiv}

\usepackage[utf8]{inputenc} 
\usepackage[T1]{fontenc}    
\usepackage{hyperref}       
\usepackage{url}            
\usepackage{booktabs}       
\usepackage{amsfonts}       
\usepackage{nicefrac}       
\usepackage{microtype}      
\usepackage{lipsum}
\usepackage{graphicx,verbatim,float,subfigure,graphicx,color,siunitx}
\usepackage{amsmath}
\usepackage[dvipsnames]{xcolor}
\usepackage{makecell}
\graphicspath{ {./images/} }
\usepackage{amssymb}
\usepackage{svg}
\usepackage{multirow}  
\usepackage[export]{adjustbox}
\usepackage{soul}
\usepackage{textcomp}
\usepackage{multirow}
\usepackage{enumitem}
\usepackage{amsfonts}
\usepackage{bbm}
\usepackage{amsthm}
\newlist{Properties}{enumerate}{2}
\setlist[Properties]{label=Property \arabic*., font=\textbf, itemindent=*}

\usepackage{epstopdf}
\AppendGraphicsExtensions{.tif}

\title{\textit{Ar$\chi$i-Textile} Composites: Role of Weave Architecture on Mode-I Fracture Energy in Woven Composites}

\author{
 Hridyesh Tewani \\
  Dept. of Civil \& Env. Engineering \\
  University of Wisconsin-Madison \\
  Madison, WI 53706 \\
   \And
  Jackson Cyvas \\
  Dept. of Mechanical Engineering \\
  University of Wisconsin-Madison \\
  Madison, WI 53706
  \And
  Kennedy Perez \\
  Dept. of Civil \& Env. Engineering \\
  University of Wisconsin-Madison \\
  Madison, WI 53706
  \And
  Pavana Prabhakar \\
  Dept. of Mechanical Engineering \\
  Dept. of Civil \& Env. Engineering \\
  University of Wisconsin-Madison \\
  Madison, WI 53706 \\
  \texttt{pavana.prabhakar@wisc.edu} \\
}

\begin{document}
\maketitle


\begin{abstract}

This paper investigates the impact of weave architectures on the mechanics of crack propagation in fiber-reinforced woven polymer composites under quasi-static loading. Woven composites consist of fabrics/textiles containing fibers interwoven at 0 degrees (warp) and 90 degrees (weft) bound by a polymer matrix. The mechanical properties can be tuned by weaving fiber bundles with single or multiple materials in various patterns or architectures. Although the effects of uniform weave architectures, like plain, twill, satin, etc. on in-plane modulus and fracture energy have been studied, the influence of patterned weaves consisting of a combination of sub-patterns, that is, architected weaves, on these behaviors is not understood. We focus on identifying the mechanisms affecting crack path tortuosity and propagation rate in composites with architected woven textiles containing various sub-patterns, hence, \textit{Ar$\chi$i} {\bf(ar.kee)} \textit{-Textile} Composites. Through compact tension tests, we determine how architected weave patterns compared to uniform weaves influence mode-I fracture energy of woven composites due to interactions of different failure modes. Results show that fracture energy increases at transition regions between sub-patterns in architected weave composites, with more tortuous crack propagation and higher resistance to crack growth than uniform weave composites. We also introduce three geometrical parameters — transition, area, and skewness factors — to characterize sub-patterns and their effects on in-plane fracture energy. This knowledge can be exploited to design and fabricate safer lightweight structures for marine and aerospace sectors with enhanced damage tolerance under extreme loads.

\end{abstract}


\section{Introduction}\label{intro}

Fiber-reinforced polymer composites have become integral materials across industries such as aerospace, automotive, and marine, owing to their superior mechanical properties while being lightweight. These composites can be tailored to achieve desired mechanical performance, boosting their functionality. Broadly, these composites are widely categorized into woven and non-woven/unidirectional composites. Compared to traditional unidirectional (UD) continuous fiber-reinforced composites, woven composites offer the added advantage of tuning the mechanical properties through hybridization, like weaving patterns with single or multiple materials. Woven composites comprise a fabric with fibers interlaced in 0-degree (warp) and 90-degree (weft) orientations \cite{onal2007modeling}. This mechanical interlocking of fibers in two orthogonal directions in various weave patterns potentially leads to complex and interacting damage mechanisms, including matrix cracking in transverse yarns, delamination in areas of interlacement, fiber pull-out, and fiber fracture \cite{lisle2015damage}. Hence, it is of utmost importance to understand the role of weave patterns, especially, architected ones, on the mechanical properties and their damage and failure mechanics. \textbf{To that end, in this study, we aim to elucidate the effect of weave patterns on tensile properties and crack propagation in uniform and architected weave composites}. 

In recent years, woven composites have gained considerable attention for their excellent strength-to-weight ratio, tailorable mechanical performance, and tunable drapability \cite{Bakar2013,ullah2015characterisation,erol2017effects,zhou2018multi,vaidya2023performance}. Previously, Ishikawa and Chou \cite{ishikawa1982,ishikawa1982elastic, ishikawa1983one} introduced various analytical models to predict the tensile properties of hybrid and non-hybrid woven composites. In their models, they accounted for fiber undulation and bridging effects near the undulation region to establish a relationship between weave architectures and the tensile properties of woven composites. Naik and Ganesh \cite{naik1995analytical} further contributed with an analytical method for thermo-elastic analysis of plain weave composites by considering yarn undulation and continuity in both warp and weft directions. Moreover, Scida et al. \cite{scida1998prediction} developed a micromechanical model, \textit{MESOTEX}, to predict the elastic properties of hybrid and non-hybrid weave composites. They applied classical lamination theory to sections of woven structure and considered the effect of undulations in warp and weft directions. Their findings showed a strong agreement between analytical and experimental results. To predict the properties of woven composites under complex in-plane stresses, Qian et al. \cite{qian2023numerical} proposed a numerical technique. They concluded that the properties of plain weave composites were inferior to those of twill- and satin-weave composites due to more undulations in plain weave. Recent research has also focused on the mechanical properties of woven composites with non-traditional weave architectures. Feng et al. \cite{Feng2022} introduced a physics-constrained neural network designed to predict the in-plane moduli values of composites with various weave architectures, including different patterns and materials. Moreover, they incorporated a backward prediction module to obtain a weave pattern or material sequence needed to achieve the desired target moduli values. 

Despite extensive studies on the effects of uniform weave architectures on in-plane moduli and fracture of woven composites, \textbf{there is a gap in understanding the impact of patterned weave architectures on fracture propagation mechanisms and the associated fracture energy} \cite{Xu2017,murdani2009fatigue}. Katafiasz et al. \cite{Katafiasz2019} proposed various modified compact tension geometries to prevent premature buckling and compressive failure during crack propagation and result in stable crack growth. To further explore the influence of sample geometry on the fracture toughness of woven composites, Dalli et al. \cite{Dalli2019} performed an experimental and numerical study on Compact Tension (CT) and Double Edge Notch Tension (DENT) specimens. They found that the crack tip radius did not impact fracture toughness, and the toughness values evaluated for both specimens were in good agreement. Cheng et al. \cite{cheng2022study} experimentally investigated the intralaminar crack propagation in plain weave carbon/epoxy composites to analyze the effect of crack tip location on the propagation. It was found that the crack path was independent of the position of the crack tip location, and the propagation was divided into two stages. Stage A corresponded to maximum energy accumulation, whereas Stage B exhibited a gradual release of energy due to fiber bundle breakage. Danodan et al. \cite{donadon2007intralaminar} performed experimental and numerical investigation of mode-I fracture toughness of unbalanced hybrid plain weave composites with 12K carbon fibers in the warp direction and S150 glass fibers in the weft. The disparity in yarn materials and sizes, in warp and weft, created an unbalanced fabric, leading to different fracture toughness in both orthogonal directions. The study concluded that the fracture mechanisms in woven composites were more complex than the UD counterpart due to interactions between yarns in the warp and weft directions and the matrix. Additionally, these interactions might introduce a local coupling between mode-I/mode-II fracture propagation as a result of a more tortuous crack path than UD laminates. They also proposed a modified correction function to consider the geometric and orthotropic effects in woven composites. Additionally, Rokbi et al. \cite{rokbi2011} studied crack propagation in glass/epoxy twill composites, and two stages were identified for crack propagation. They speculated that the first load drop was due to matrix cracking and fiber-matrix debonding. Following the first load drop, the predominant failure mechanism was fiber breakage. Boyina et al. \cite{boyina2014mixed} characterized the translaminar fracture toughness of plain weave composites under mixed-mode loading. It was shown that a higher ratio of Mode-I to Mode-II resulted in more fiber-matrix debonding as opposed to fiber breakage, lowering the fracture toughness for crack initiation. Liu and Hughes \cite{liu2008fracture} experimentally investigated the fracture toughness of woven flax-fabric/epoxy composites with varying yarn density, weave patterns, and stacking sequence.  They found that the toughness was dominated by the volume fraction of the reinforcement (varying weft yarn density) and the direction of crack propagation. This dependence on directionality was attributed to differences in tensile strengths in warp and weft directions. They also concluded that the toughness was not affected by the weave architectures for a particular yarn density. Blanco et al. \cite{Blanco2014} studied the fracture toughness of composites with a uniform weave pattern, 5-H satin, with two configurations - crack direction parallel and perpendicular to the warp direction. Their study observed tow-splitting with the associated damage when the warp was perpendicular to the crack direction because of longer carbon fiber floats along that direction. 

While prior studies have demonstrated that weave architectures result in complex fracture mechanisms in woven composites, they are limited to investigating uniform weave patterns. The current work aims to experimentally \textbf{establish a relationship between weave architectures and the complex mechanics of crack propagation within woven composites}. By elucidating the underlying mechanisms governing crack growth in weave patterns, this study \textbf{provides invaluable insights into the design and optimization of composites with modified weave architectures for specific applications}.

\section{Motivation}\label{motiv}

In this paper, we investigated the Mode-I intralaminar fracture energy in architected woven composites to establish how the weave architectures dictate the mechanics of crack propagation. First, we selected six weave patterns - three uniform and three architected - and subjected them to tensile loading. Although the tensile properties of traditional woven composites have been studied in depth, architected woven composites can be further tailored to achieve the desired mechanical performance. After assessing the impact of weave architectures on tensile properties, we performed Mode-I compact tension tests to explore the relationship between weave architectures and fracture energy. We then proposed three geometric parameters for architected weave patterns to design woven composites with improved damage tolerance and mechanical performance. 

Specifically, the overarching goal of this current work is to answer the following questions:
\begin{itemize}
    \item How do different weave architectures influence the tensile performance of woven composites? 
    \item Can introducing architectures in weave patterns enhance the damage resistance of woven composites?
    \item What geometrical features of weave architectures influence the tensile behavior and fracture energy of woven composites? 
\end{itemize}

\section{Methodology}\label{method}
This section discusses the weave patterns considered in the current study and the manufacturing process of woven composites. Then, we describe the test methods to evaluate the tensile properties and effects of weave architectures on fracture energy.

\subsection{Materials and manufacturing}\label{mat}

\begin{figure}[h!]
\centering
\includegraphics[width=0.98\textwidth]{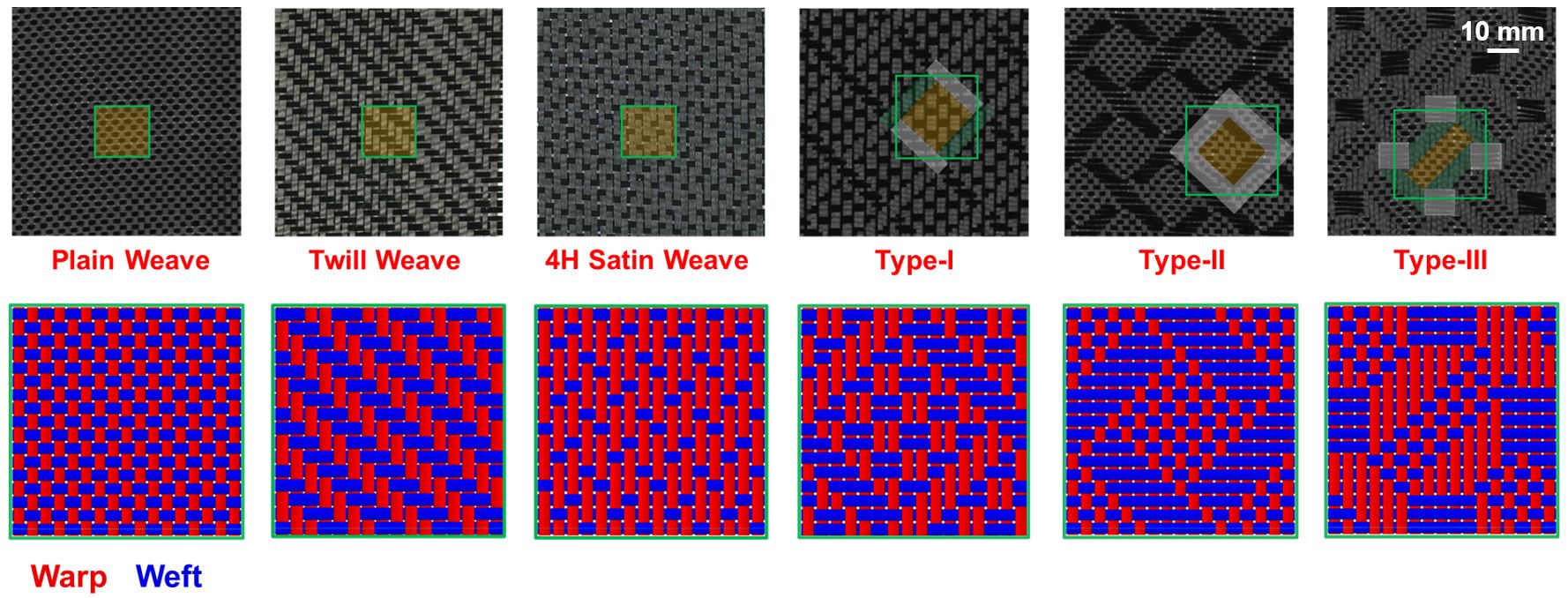}
\caption{Uniform and architected weave patterns with their 16x16 unit cells are considered in this study. For the architected weave patterns, the areas of sub-patterns within the representative unit cell (RUC) are highlighted.}
\label{img:weavepatt}
\end{figure}

We procured five carbon fiber fabrics from Fibre Glast Corp. and a 4-Harness satin fabric from Composite Envisions LLC to manufacture composite laminates. Table \ref{tab:mat_weave} summarizes the weave parameters for all six weave patterns. We selected three uniform and three architected weave patterns; all six patterns are shown in \textbf{Figure} \ref{img:weavepatt}. The architected weaves consist of sub-patterns within patterns, as highlighted in \textbf{Figure} \ref{img:weavepatt}. These sub-patterns are uniform weave patterns arranged in specific shapes and sizes within the representative unit cell (RUC) of architected weave patterns. In Type-I weave, the design consists of a basket weave in the center, surrounded by a 3x3 twill weave. On the other hand, Type-II and Type-III weaves comprise 5-Harness satin weave surrounding plain weave in the center, with each type having a different arrangement. In the current work, composites containing woven textiles with sub-patterns within patterns are termed as \textit{Ar$\chi$i-Textile} Composites, where Ar$\chi$i stands for architected). 

\begin{table}[h!]
\centering
\renewcommand{\arraystretch}{1.2}
\caption{Properties and nomenclature of the carbon fiber fabrics considered in this study \cite{Plain, Twill, MBoss, Diam, Rosw}} 
\resizebox{\columnwidth}{!}{
\begin{tabular}{lccccc}
\hline
    \textbf{Weave pattern} & \textbf{Yarn thickness (Warp)} & \textbf{Yarn thickness (Weft)}  & \textbf{Areal weight (oz/sq yd)} & \textbf{Group} & \textbf{Nomenclature}   \\ \hline
    
    \textit{Plain weave} & 3K & 3K & 5.4 - 5.9 & Uniform & PW   \\ 
    
    \textit{Twill weave} & 3K & 3K & 5.7 - 5.9 & Uniform & TW   \\ 

    \textit{4-H Satin weave} & 3K & 3K & 6.4 & Uniform & 4HSW   \\ 
    
    \textit{M-Boss weave} & 3K & 3K & 6 & Architected & Type - I   \\ 

    \textit{Diamond-plate weave} & 3K & 3K & 6 & Architected & Type - II   \\ 

    \textit{Roswell weave} & 3K & 3K & 6 & Architected & Type - III   \\ \hline

\end{tabular}
}
\label{tab:mat_weave}
\end{table}

We chose West System 105 epoxy resin with West System 209 extra-slow hardener in a 3:1 ratio (by volume) to manufacture woven composite. We used a hand-layup process with 12 layers of woven fabric to manufacture a carbon fiber laminate with a thickness of $\approx$ 4mm. To avoid the creation of nesting, each layer of fabric was carefully placed to align all unit cells during the hand layup process. It is important to ensure this, as significant nesting can alter the composite behavior.

\subsection{Overview of existing weave parameters} \label{met:weavechar}
\begin{figure}[h!]
\centering
\includegraphics[width=0.98\textwidth]{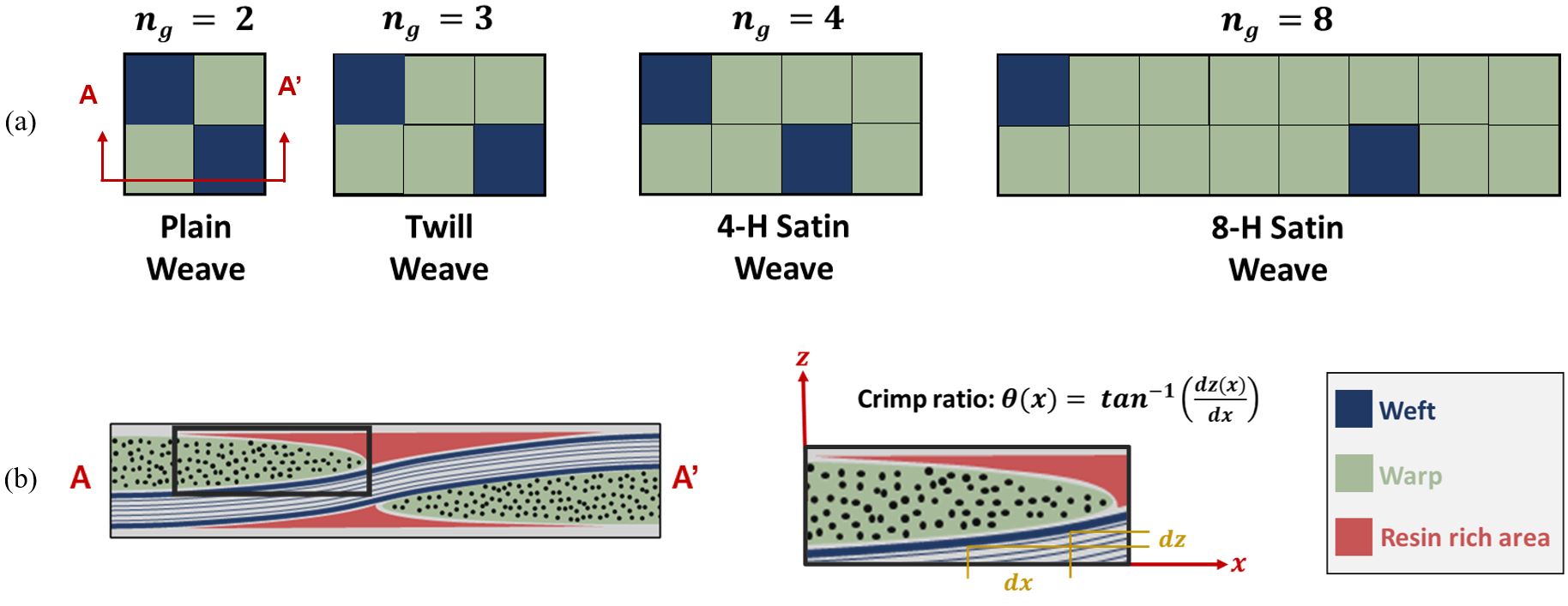}
\caption{(a) $n_{g}$ values for Plain, Twill, 4-H Satin, and 8-H Satin weave patterns; and (b) Section AA' of the plain weave to illustrate crimp ratio at the interlaced region \cite{Feng2022}. (\textit{Image reused with permission})}
\label{img:WeavePar}
\end{figure}

Various parameters have been suggested in previous studies to characterize uniform weave patterns. These parameters include \textit{$n_{g}$}, \textit{crimp ratio}, and \textit{yarn mobility} \cite{ishikawa1982,osada2003initial,Tapie2016}. The geometrical parameter, $n_{g}$ is defined as the number of warp yarns intertwined with a single weft yarn. Another geometrical parameter, crimp ratio ($\theta (x)$), defines the degree of undulation at the interlaced area. A lower $n_{g}$ value for a woven fabric results in a higher crimp ratio, leading to inferior mechanical properties. Osada et al. \cite{osada2003initial} showed that the 5-H satin weave composite ($n_{g}$=5) displayed significantly superior tensile modulus and strength compared to plain weave composite ($n_{g}$=2). The $n_{g}$ and crimp ratio are illustrated in \textbf{Figure} \ref{img:WeavePar}. Finally, yarn mobility is a factor that is defined as the ease of yarn movement in a given woven fabric \cite{Tapie2016}. A weave with a lower $n_{g}$ value has lower yarn mobility due to friction at more areas of undulations. The yarn mobility can alter the failure process from fiber breakage in woven composites with lower yarn mobility as opposed to fiber pull-out in woven composites with higher yarn mobility. 

Although these parameters have been used to establish a relationship between the weave architectures and the mechanical properties, they are limited to characterizing uniform weave patterns. Previously, Feng et al. \cite{Feng2022} proposed using Gray Level Co-occurrence Matrices (GLCM) to extract geometrical parameters to describe architected weave patterns. GLCM is a statistical texture-classification method that is used to analyze spatial relationship of pixels with specific gray levels in a given image \cite{haralick1973textural}. These parameters were then related to the in-plane properties to optimize weave patterns. In their study, they represented weave patterns as a matrix of '0' and '1' binary values in which '1' indicated warp below weft and '0' indicated warp above weft. Since the weave pattern was classified as a binary matrix, the gray levels were '0' and '1'. Consequently, GLCM matrices in different directions (vertical, horizontal, diagonal) were 2-by-2 matrices derived from transitions between $0\rightarrow0$, $0\rightarrow1$, $1\rightarrow0$, and $1\rightarrow1$. For instance, a 4x4 plain weave unit cell can be represented as $\begin{bmatrix}
    1 & 0 & 1 & 0 \\
    0 & 1 & 0 & 1 \\
    1 & 0 & 1 & 0 \\
    0 & 1 & 0 & 1 
\end{bmatrix}$. The GLCM matrices in both vertical and horizontal directions would be $\begin{bmatrix}
    0 & 6 \\
    6 & 0
\end{bmatrix}$. This indicates that in a plain weave, there are only inhomogeneous ($0\rightarrow1$ and $1\rightarrow0$ shown as the off-diagonal terms of the GLCM matrix) transitions. In this work, we use GLCM matrices to obtain modified geometrical parameters to represent architected weave patterns in Section \ref{res:weavepar}.

\subsection{Mechanical testing}
\subsubsection{Tensile testing}\label{met:char}
We performed uniaxial tensile tests using the MTS Criterion 43 test frame with a load cell capacity of 50 kN to assess the influence of weave architectures on mechanical response. The samples were waterjet cut with dimensions of 250mm x 35mm to ensure the inclusion of at least 1 unit cell across the width. Further, to prevent stress concentration in the gripping area, we attached Garolite G-10/FR4 Sheets tabs with a length of 50mm at the ends of the sample \cite{zhou2021experimental}. We performed these tests according to ASTM D3039 \cite{ASTM3039} at a crosshead displacement rate of 2mm/minute. These tests were performed as a set of three, and we used an MTS contact extensometer with a gauge length of 50mm to accurately quantify strain. We also used a correlated solutions digital image correlation (DIC) system with a 10 Hz image capture frequency to analyze the effects of weave architectures on the full-field strain.

\subsubsection{Compact tension testing}\label{met:CT}

\begin{figure}[h!]
\centering
\includegraphics[width=0.9\textwidth]{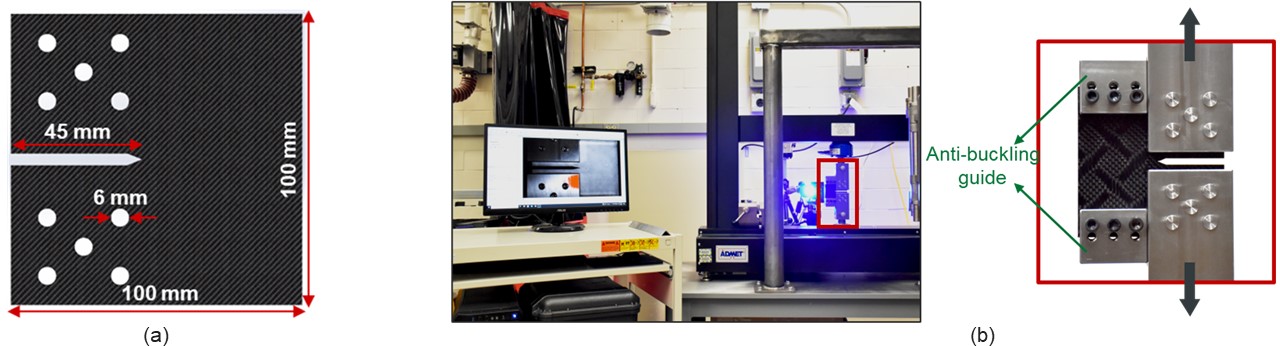}
\caption{(a) 5-pin compact tension sample geometry with 45mm crack length, and (b) Compact tension test setup used for fracture testing monitored with DIC camera.}
\label{img:setup}
\end{figure}

We performed Mode-I compact tension tests on 3 samples for each weave pattern in compliance with ASTM E399 \cite{ASTM399} on the ADMET eXpert 2613 test frame equipped with a load cell of 50 kN. We designed a modified 5-pin fixture with an anti-buckling guide to perform fracture tests; a schematic of the sample configuration is shown in \textbf{Figure} \ref{img:setup}(a) \cite{Laffan2013}. We loaded the samples at a 5 mm/minute crosshead displacement rate. In addition, we used the correlated solutions DIC system with an image capture frequency of 50 Hz and Vic-2D post-processing software to obtain the fracture process zone (FPZ) at the crack tip. The experimental setup for this test is shown in \textbf{Figure} \ref{img:setup}(b). The area method was used to obtain the fracture energy ($G_{Ic}$) associated with every crack propagation \cite{LAFFAN2010606}.

\section{Results and Discussion}\label{resdis}
In this section, we explore the geometric parameters that describe architected weave patterns, which will be used later to establish their relationship with the mechanical performance of woven composites. We will then discuss how different weave architectures influence the tensile behavior of woven composites. Finally, we will examine the role of weave architectures on the fracture energy of woven composites and discuss any correlation between geometric parameters and fracture energy.

\subsection{Architected weave parameters: understanding mechanical response}\label{res:weavepar}

\begin{figure}[h!]
\centering
\includegraphics[width=0.75\textwidth]{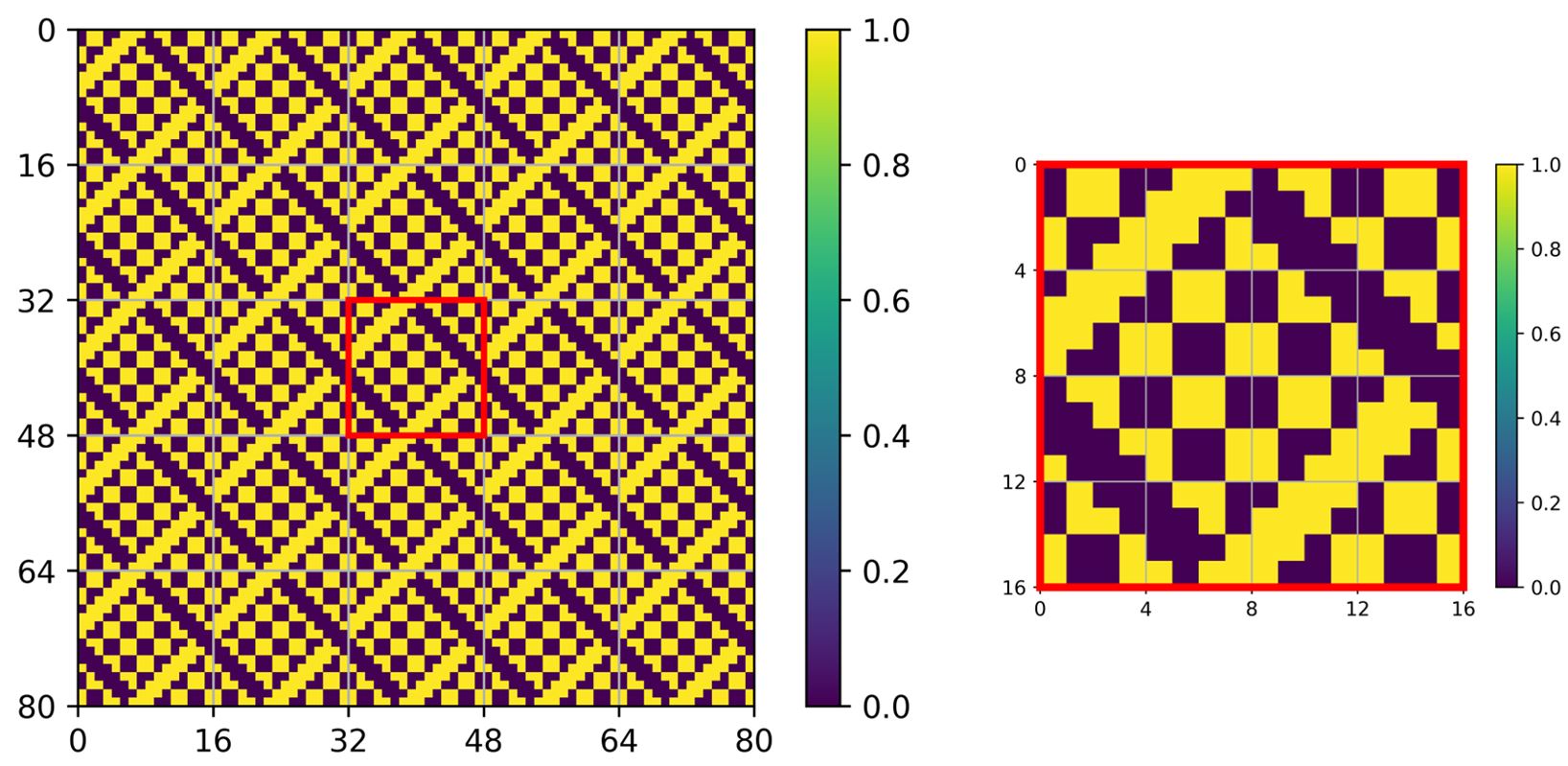}
\caption{An illustration of a unit cell extracted from a full weave pattern for Type-I weave represented as a matrix of `0' and `1' binary values. `1’ indicates warp below weft and `0’ indicates warp above weft. In this study, geometrical factors are defined for a 16x16 unit cell, though the representative unit cell size can be adjusted to any size to define these parameters.}
\label{img:RUC}
\end{figure}

Unlike uniform weaves, which can be described using existing parameters, architected weaves require a definition of modified geometrical parameters to capture sub-patterns within the overall pattern. In this section, we will discuss three geometrical factors for a unit cell of architected weaves: i) transition factor, ii) area factor, and iii) skewness factor. These factors are defined for a unit cell which is periodically repeated to form a larger weave pattern as demonstrated for Type-I weave in \textbf{Figure} \ref{img:RUC}.

We define the transition factor ($\delta$) to characterize the change between different sub-patterns. Typically, uniform weave composites with lower ${n_g}$ show fiber fracture, while those with higher ${n_g}$ exhibit both, fiber pull-out and fracture. Hence, it is critical to capture this transition between patterns and their influence on crack propagation. Thus, $\delta$ accounts for variation in ${n_g}$ at these transition regions. It should be noted that two weave patterns can have the same ${n_g}$, for instance, plain and baskets weaves. Therefore, we also need to consider the effect of different crimp ratios in the transition regions between weave patterns. The effect of different crimp ratios is quantified by inhomogeneity (IH) from the GLCM matrix in the form $\begin{bmatrix}
    a & b \\
    c & d
\end{bmatrix}$ as shown in Equation \ref{Eqn:IH}. 

A higher inhomogeneity indicates a higher crimp ratio. For instance, IH values for plain and basket weaves are 1 and 0.466, respectively. Although both weave patterns have $n_g$ = 2, the plain weave has a higher crimp ratio than the basket weave due to tighter packing. Therefore, $\delta$ can be defined between two patterns, as shown in Equation \ref{Eqn:delta}.  

\begin{figure}[h!]
\centering
\includegraphics[width=0.88\textwidth]{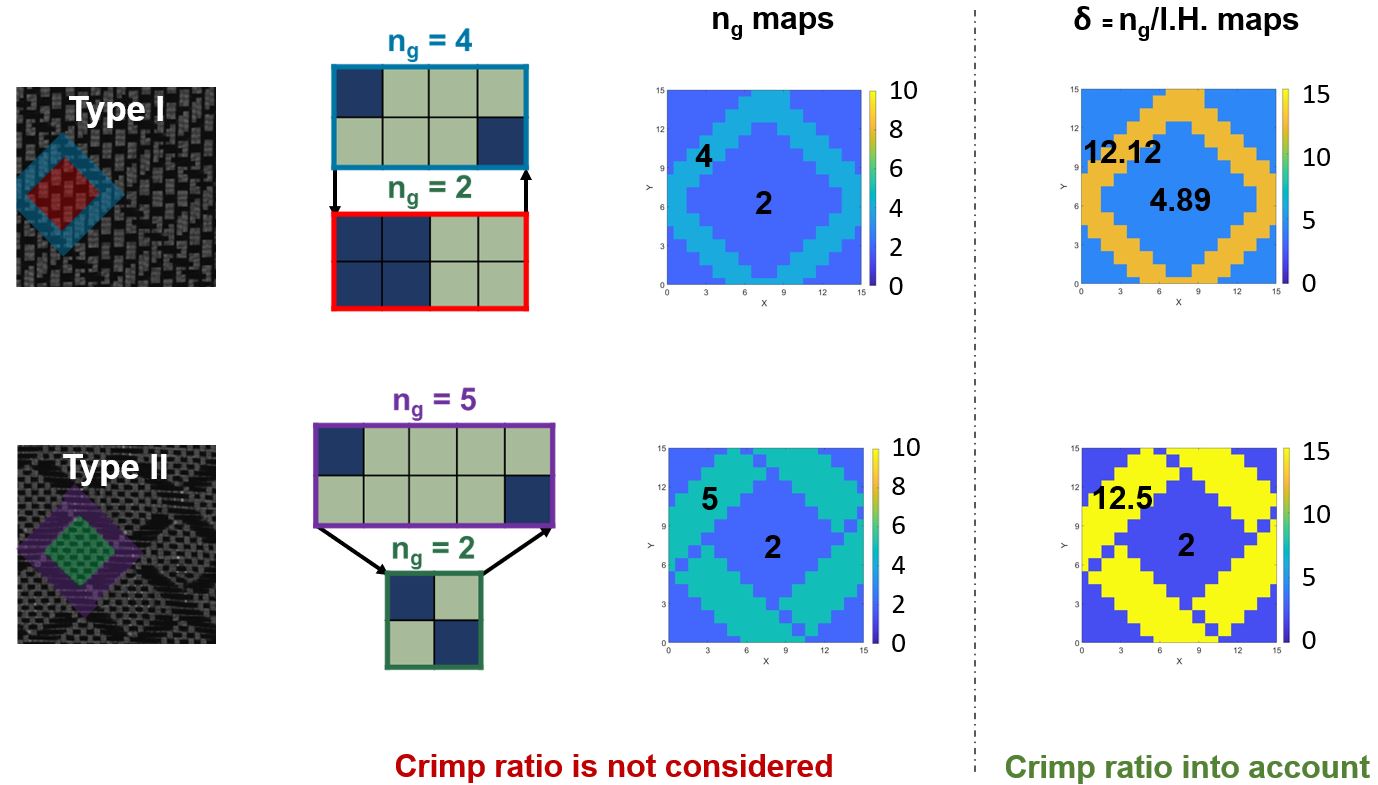}
\caption{Illustration of transition factor ($\delta$) in Type-I and Type-II weave patterns. Type-II and Type-III weave patterns have the same transition factors.}
\label{img:resDeltaFac}
\end{figure}

In \textbf{Figure} \ref{img:resDeltaFac}, we demonstrate transition factors for Type-I and Type-II weaves. In Type-I, the sub-patterns transition from $n_g = 4$ (3x3 twill) to $n_g = 2$ (basket), while in Type-II, they transition from $n_g = 5$ (5-H satin) to $n_g = 2$ (plain). The I.H. values for all sub-patterns are calculated using the GLCM method, and the $n_g$ values are divided by the I.H. values to generate $n_g / I.H.$ maps for both weaves. The transition factors ($\delta$) are then derived using Equation \ref{Eqn:delta}.

\begin{figure}[h!]
\centering
\includegraphics[width=0.8\textwidth]{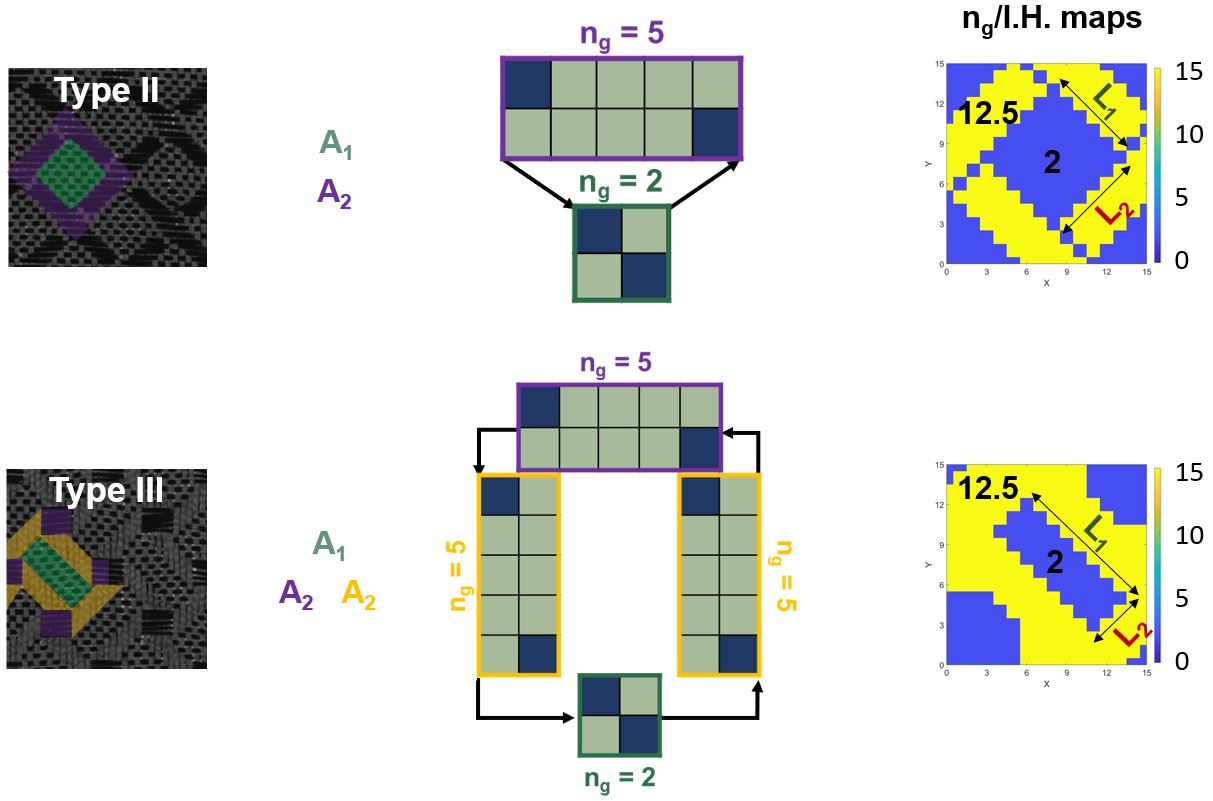}
\caption{Area ($\rho$) and Skewness ($\sigma$) factors to characterize the shape of sub-pattern with lower $n_g$ within a higher $n_g$ sub-pattern.}
\label{img:resAreaFac}
\end{figure}

However, we observe that Type-II and Type-III weaves have similar $\delta$ values as shown in \textbf{Figure} \ref{img:resAreaFac}. Hence, we further introduce area and skewness factors to characterize the size and shape of the sub-pattern within a pattern. The area factor ($\rho$) determines the ratio between the areas of sub-patterns involved in transitions as defined by Equation \ref{Eqn:rho}. Whereas, the skewness factor ($\sigma$) captures the shape governed by the distance between the centroids of two sub-patterns as shown in Equation \ref{Eqn:sigma}. We chose the three architected weaves, Type-I, Type-II, and Type-III to understand the impact of their geometrical factors. Type-I and Type-II weaves have similar $\sigma$ but different $\delta$ and $\rho$. On the other hand, Type-II and Type-III have the same $\delta$ but different $\rho$ and $\sigma$. The three geometrical factors for Type-I, Type-II, and Type-III weave composites are summarized in Table \ref{tab:weavepar}. 

In this study, the goal was not to propose a global parameter combining all three proposed factors; instead, the focus was to understand the physics behind the observed crack propagation. Establishing a global parameter would require a bigger dataset. Instead, considering each factor separately allowed us to understand their specific contributions to composites' fracture energy.

\boldmath
\begin{subequations}   
\begin{equation}
\label{Eqn:IH}
    Inhomogeneity\ (IH)\ =\ \frac{b+c}{a+b+c+d}
\end{equation}
\begin{equation}
\label{Eqn:delta}
    Transition\ factor\ (\delta)\ =\ \Bigg| {\biggl(\frac{n_g}{IH}\biggr)_{\textcolor{teal}{High\ {n_g}}}} -\quad {\biggl(\frac{n_g}{IH}\biggr)_{\textcolor{red}{Low\ {n_g}}}} \Bigg|
\end{equation}
\end{subequations}
\unboldmath

\boldmath
\begin{subequations}  
\begin{equation}
\label{Eqn:rho}
    Area\ factor\ (\rho)\ =\ \frac{Area\ of\ \textcolor{red}{lower\ {n_g}} \ ({A_1})}{Area\ of\ surrounding\ \textcolor{teal}{higher\ {n_g}} \ ({A_2})}
\end{equation}

\begin{equation}
\label{Eqn:sigma}
    Skewness\ factor\ (\sigma)\ =\ \frac{\textcolor{teal}{Longer}\ length\ of\ shape\ in\ lower\ {n_g} \ ({L_1})}{\textcolor{red}{Shorter}\ length\ of\ shape\ in\ lower\ {n_g} \ ({L_2})}
\end{equation}
\end{subequations}
\unboldmath

\subsection{Tensile performance}\label{res:Tens}
\begin{figure}[h!]
\centering
\includegraphics[width=0.4\textwidth]{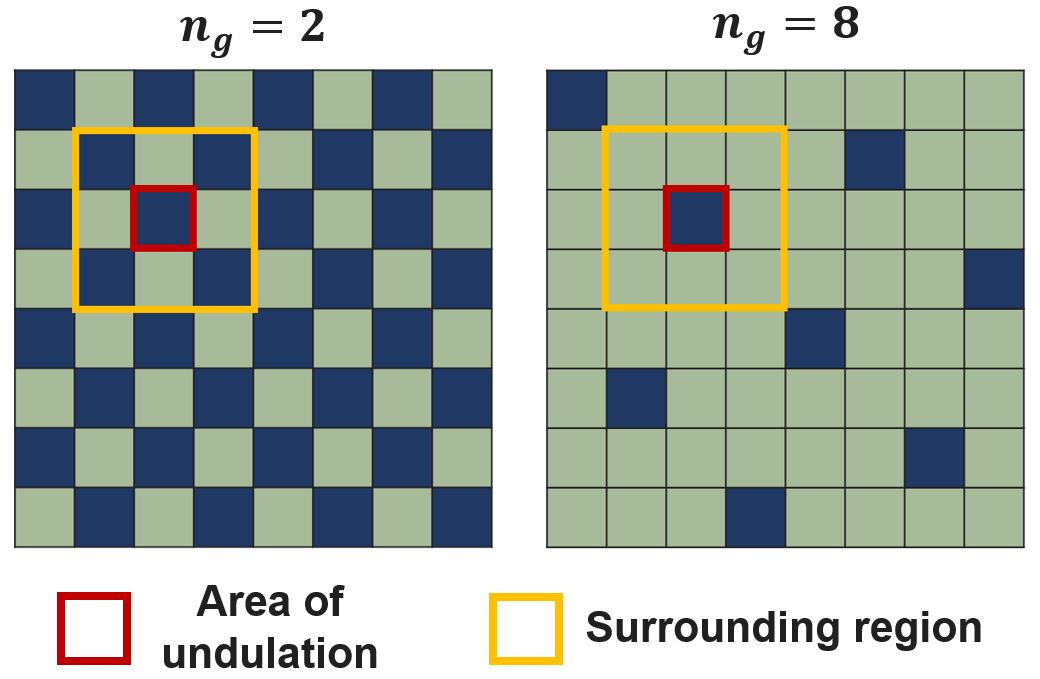}
\caption{Illustration of the surrounding region around the area of undulation for plain and 8-H satin weaves. In the 8-H satin weave, the surrounding region can be approximated as a cross-ply composite, functioning as an effective "load-carrying bridge" to distribute stress concentration at the area of undulation \cite{Feng2022}. (\textit{Image reused with permission})}
\label{img:resbridgingmod}
\end{figure}

\begin{figure}[h!]
\centering
\subfigure[]{
\includegraphics[width=0.4\textwidth]{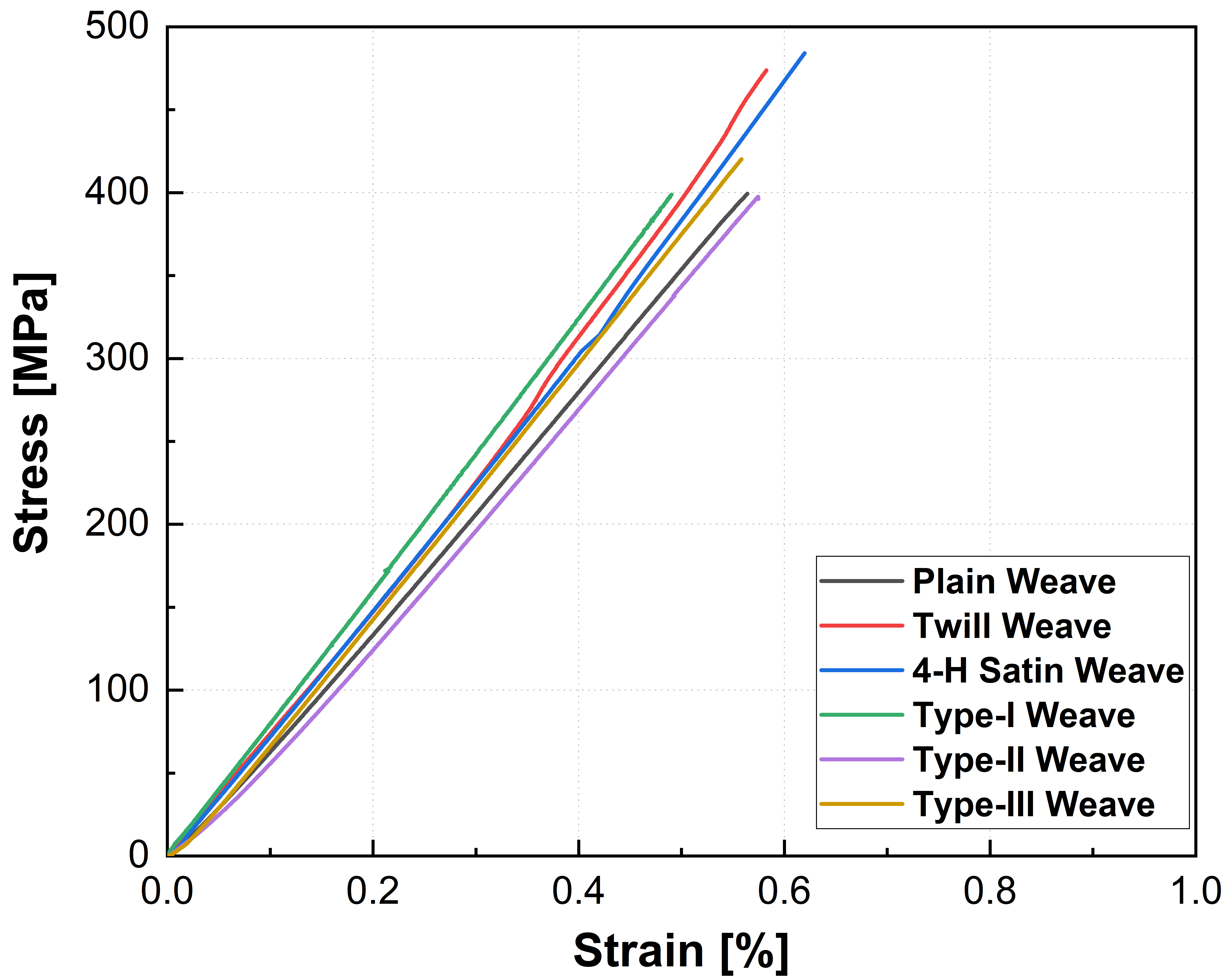}
\label{img:resten1}
}
\subfigure[]{
\includegraphics[width=0.43\textwidth]{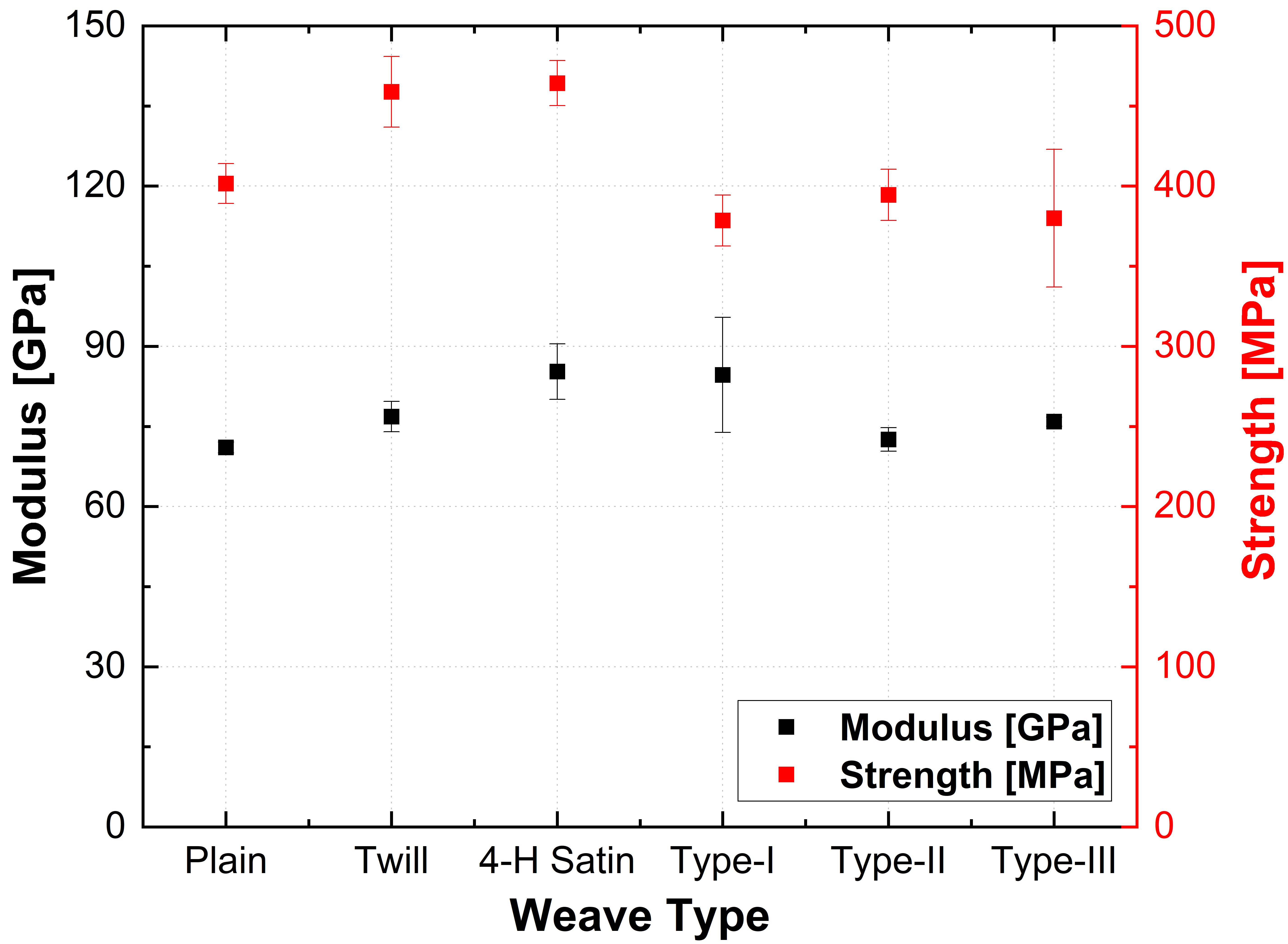}
\label{img:resten2}
}
\caption{(a) Representative tensile stress-strain plots illustrating the mechanical behavior of woven composites, and (b) Summarized tensile properties for all uniform and architected weave composites.}
\end{figure}

We first compare the tensile properties between the three uniform and three architected weave composites. \textbf{Figure} \ref{img:resten1} shows the representative graphs of the stress-strain responses of all weave architectures considered in this study and the corresponding tensile properties are shown in \textbf{Figure} \ref{img:resten2}. As discussed in Section \ref{met:weavechar}, lower ${n_g}$ implies more undulations and a higher crimp ratio, leading to lower modulus. Additionally, lower ${n_g}$ reduces the strength due to inefficient stress transfer between the areas of undulations. The "bridging model" proposed by Ishikawa and Chou \cite{ishikawa1982} considered the areas around undulations as load-carrying bridges in woven composites with ${n_g} \geq 4$. An illustration of surrounding regions for plain and 8-H satin weaves is depicted in \textbf{Figure} \ref{img:resbridgingmod}. A better stress distribution around the area of undulation in woven composites will result in higher strength. 

Among the uniform weave composites, we observe that the plain weave exhibited the lowest strength and modulus compared to the twill and satin weaves. This behavior is attributed to the lowest ${n_g}$ = 2 and the highest crimp ratio in plain weave. We also observed that satin weave composites exhibited higher strength and modulus than twill weave composites. However, the increase in properties was more significant from plain weave to twill weave composites. In a previous study by Qian et al. \cite{qian2023numerical}, similar behavior was observed for uniform weave composites. Our results show that the tensile properties will plateau with increasing values of ${n_g}$ for uniform weave composites. 

In the case of architected weave composites, Type-I weave composites showed a higher modulus in the same range as the satin weave composites. However, the strength was lower and closer to that of plain weave composites. The higher modulus is attributed to higher ${n_g}$ values and lower crimp ratios associated with the 3x3 twill and basket sub-patterns within the Type-I weave composites. We speculate the lower strength is due to areas of discontinuities or jumps in $n_g$ values between sub-patterns. For Type-II and Type-III weave composites, we noticed that the moduli and strength were in the same range as the plain weave composites. This behavior is attributed to plain weave sub-pattern regions in Type-II and Type-III weaves. These regions of plain weave dictated the loading response and failure, which limited the tensile properties of these composites, even with inclusions of sub-patterns with high ${n_g}$. 

\begin{figure}[h!]
\centering
\includegraphics[width=0.98\textwidth]{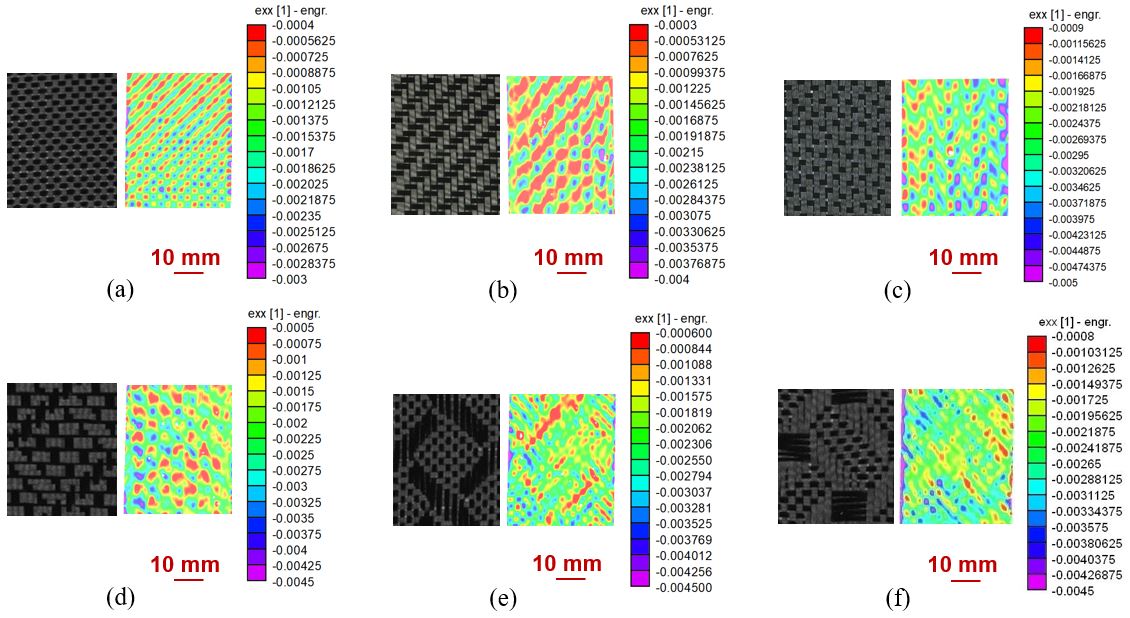}
\caption{Strain maps in the x-direction (${\epsilon_{xx}}$) on tensile samples at a global strain of 0.4\% in the y-direction (${\epsilon_{yy}}$ = 0.4\%), illustrating the impact of weave patterns on strain concentrations. The transition factor ($\delta$) influences stress concentration formation in architected weave composites. Lower $\delta$ in Type-I composites show distributed concentration areas, whereas higher $\delta$ in Type-II and Type-III composites reveal more localized areas.}
\label{img:resTenDIC/}
\end{figure}

The analysis of strain maps in the transverse direction (${\epsilon_{xx}}$) to loading for all composites at a global strain (${\epsilon_{yy}}$) of 0.4\% in the y-direction are presented in \textbf{Figure} \ref{img:resTenDIC/}. These strain maps reveal the relationship between weave patterns and strain distribution under tensile loading. Firstly, for uniform weave composites shown in \textbf{Figure} \ref{img:resTenDIC/}(a-c), strain concentrations appear in the weft yarns. This occurs due to stress concentrations at the regions of undulations where the warp and weft interlace, and because the wefts are on top of the back surface where the strain maps were captured. Consequently, in plain weave and twill weave composites, we observe continuous lines at an angle, while the 4-H Satin weave composites have discrete locations for strain concentrations. These locations correspond to weft yarns on top in these weave patterns. On the other hand, for architected weave composites (\textbf{Figure} \ref{img:resTenDIC/}(d-f)), the behavior is influenced by the transition factor $\delta$ of the weave pattern. For Type-I weave composites with a smaller $\delta$, we observed strain maps exhibited more uniformly spread strain concentration areas at the location of weft yarns. In contrast, Type-II and Type-III weave composites, with higher $\delta$, displayed strain concentrations at the transition regions between two sub-patterns. This occurs because higher $\delta$ implies different loading responses in sub-patterns, leading to stress concentrations at transitions. These different responses occur due to different ${n_g}$ and crimp ratios in sub-patterns, like, plain and 5-H satin sub-patterns in Type-II and Type-III weave composites. Conversely, lower $\delta$ results in more uniform loading across sub-patterns due to similar $n_g$ and crimp ratio, like basket and 3x3 twill sub-patterns in Type-I weave composites. 

The findings indicate that the weakest sub-pattern, that is, with the lowest ${n_g}$, determines the tensile modulus and strength of architected weave composites. Further, the $\delta$ factor is crucial to design patterns for the effective distribution of stress under tensile loading. Therefore, strategically combining sub-patterns to form architected patterns can thus tune the tensile performance and behavior of woven composites. 

\subsection{Fracture response}
\subsubsection{Crack growth analysis}\label{res:CrGr}
\begin{figure}[h!]
\centering
\subfigure[]{
\includegraphics[width=0.31\textwidth]{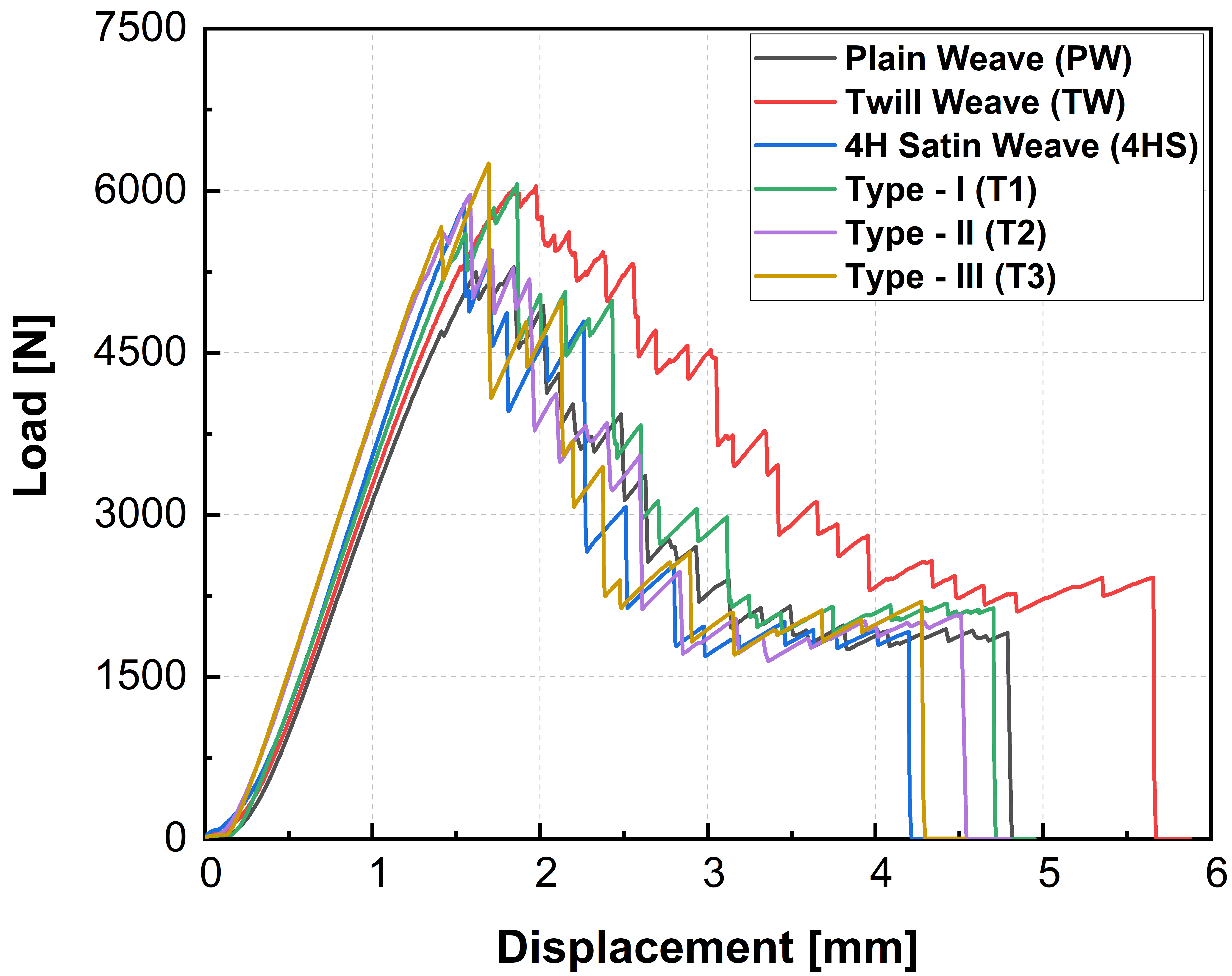}
\label{img:resfractLvD}
}
\subfigure[]{
\includegraphics[width=0.31\textwidth]{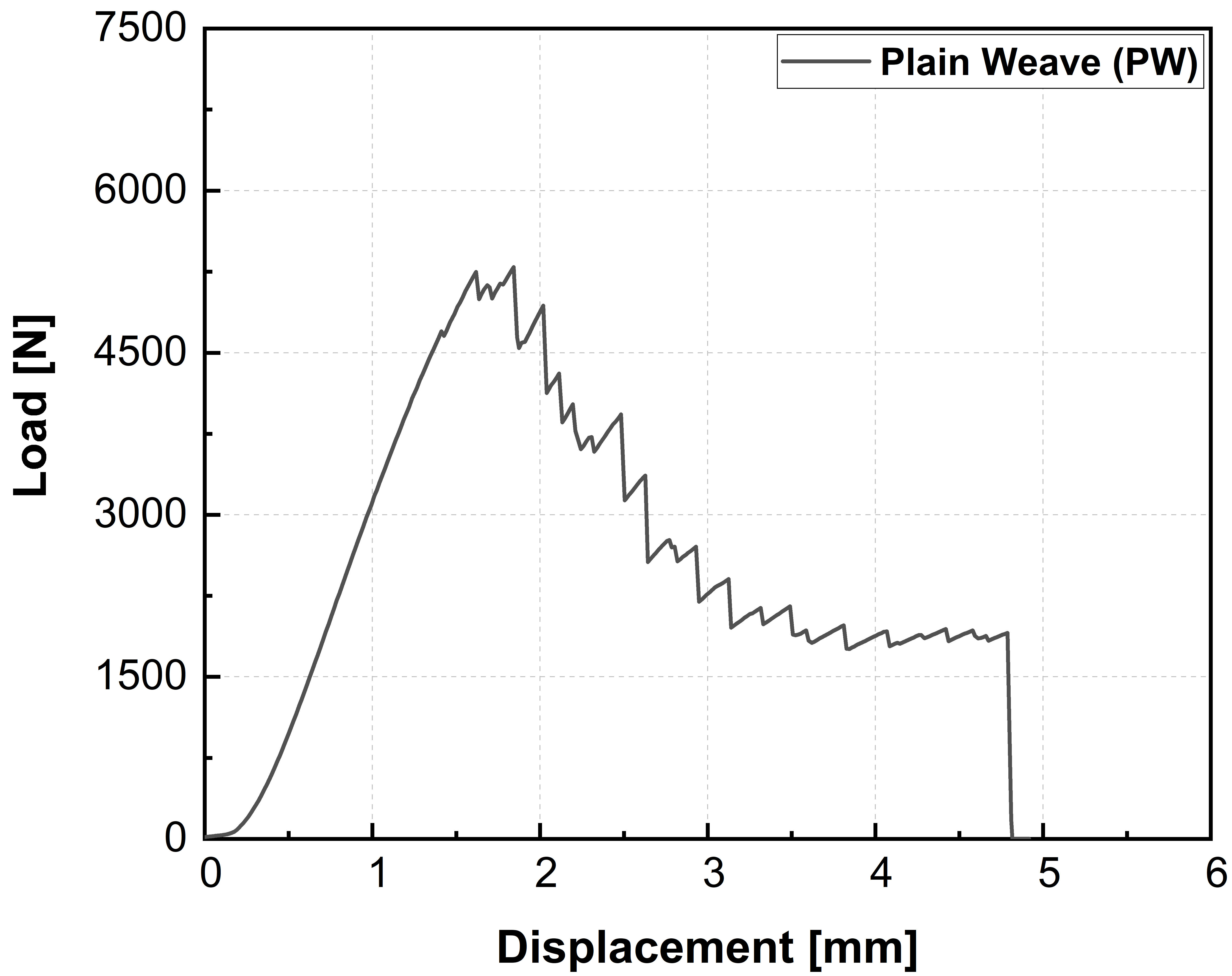}
\label{img:resfractLvDPW}
}
\subfigure[]{
\includegraphics[width=0.31\textwidth]{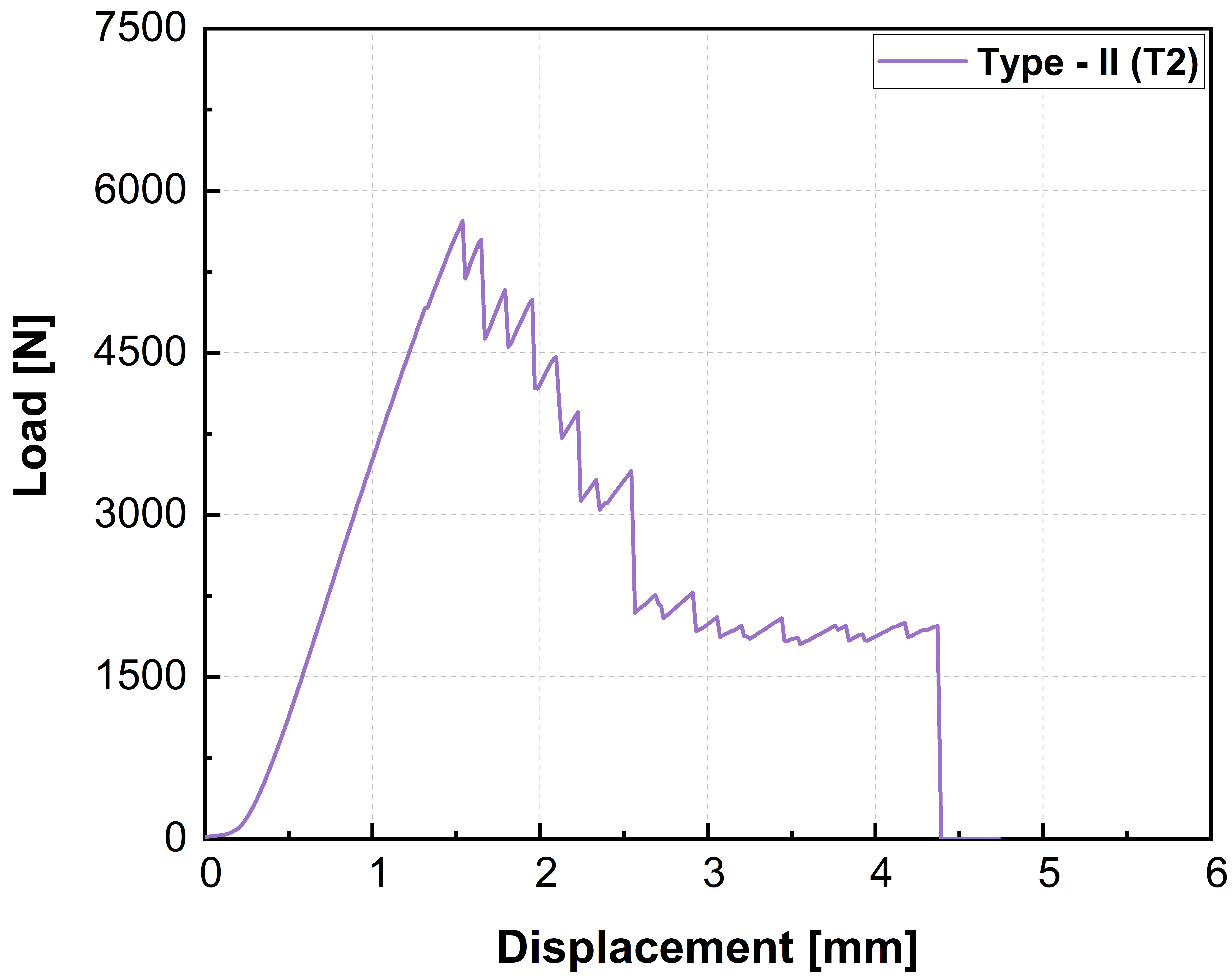}
\label{img:resfractLvDT2}
}
\subfigure[]{
\includegraphics[width=0.3\textwidth]{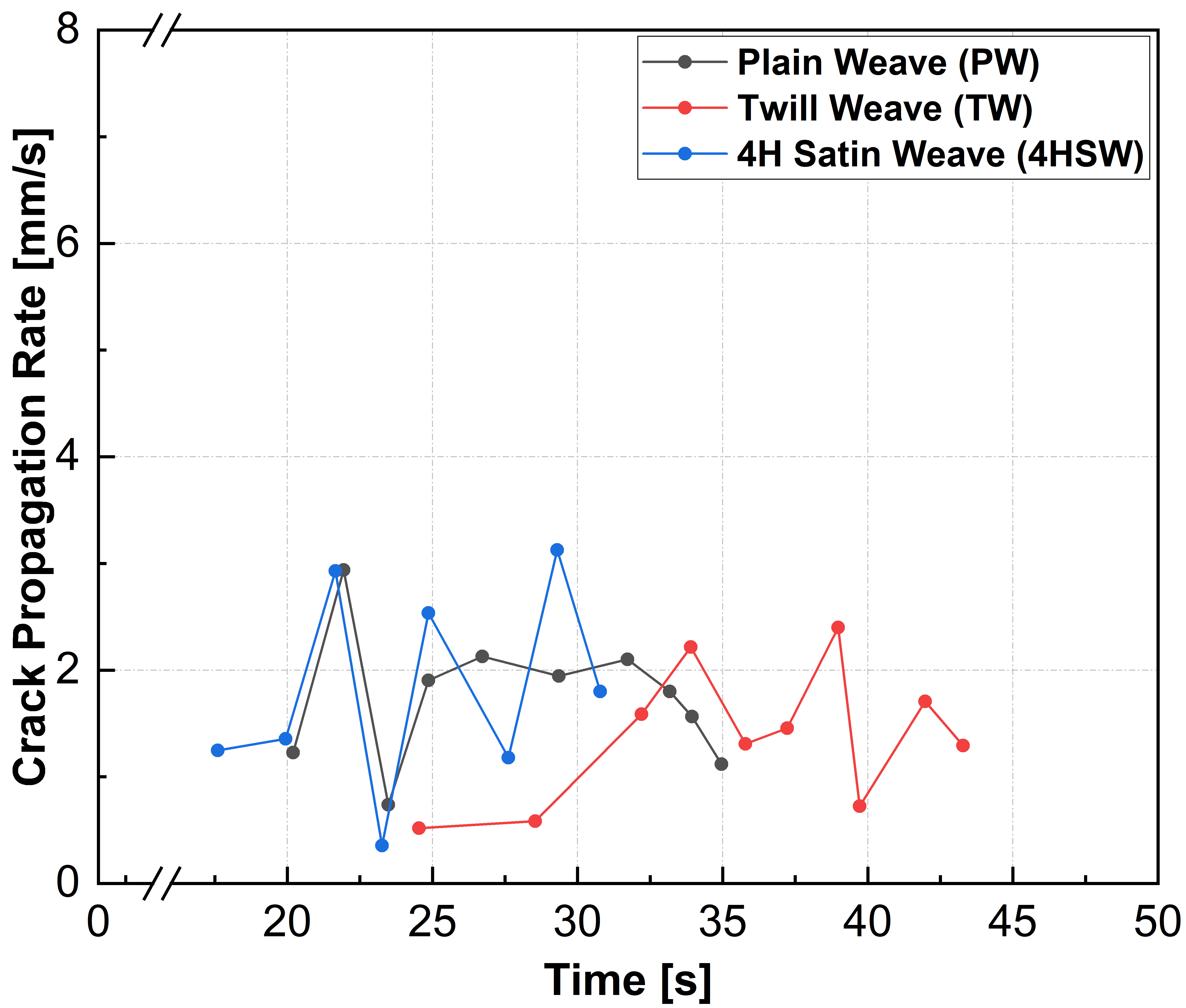}
\label{img:ratevstime1}
}
\subfigure[]{
\includegraphics[width=0.3\textwidth]{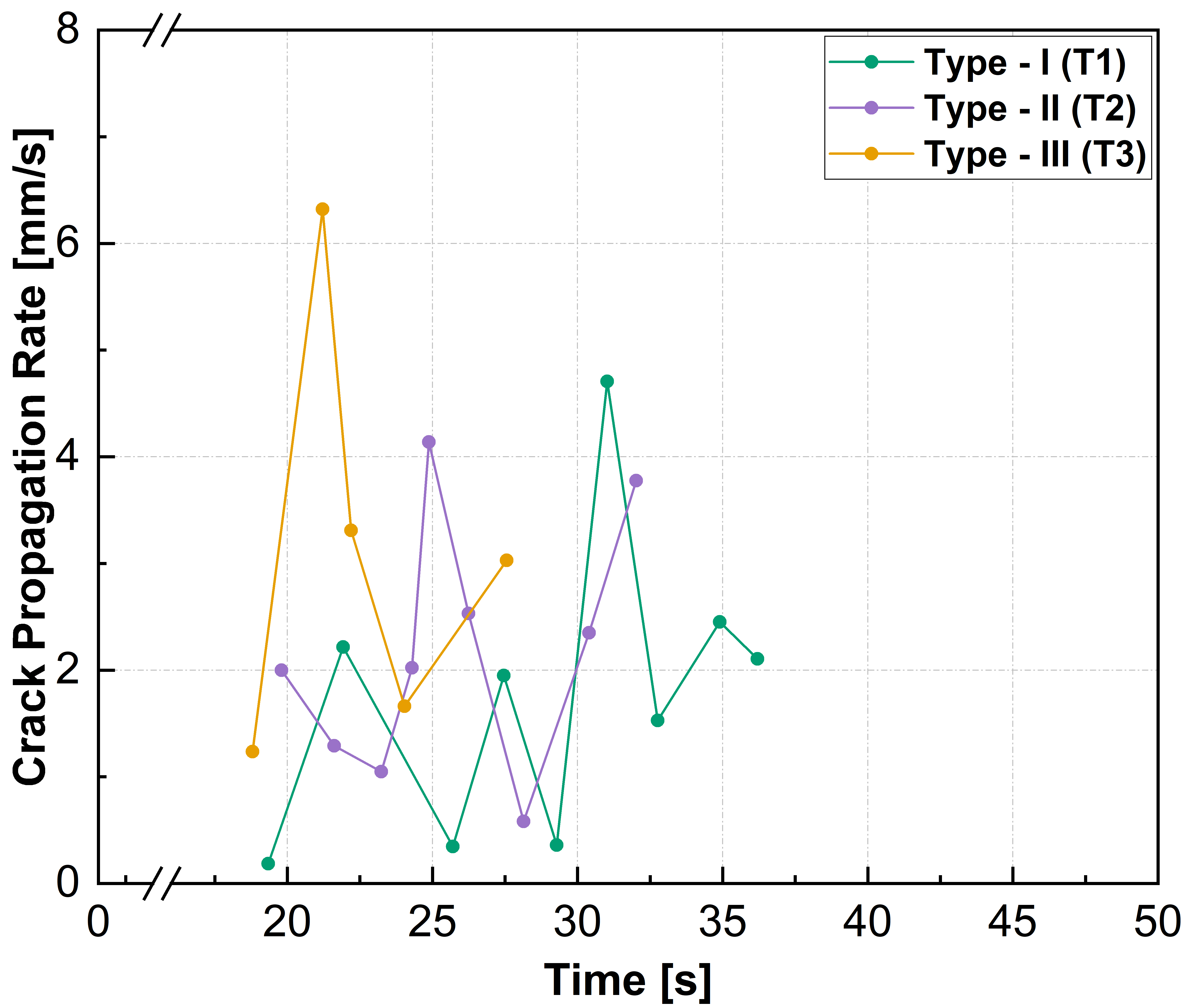}
\label{img:ratevstime2}
}
\subfigure[]{
\includegraphics[width=0.32\textwidth]{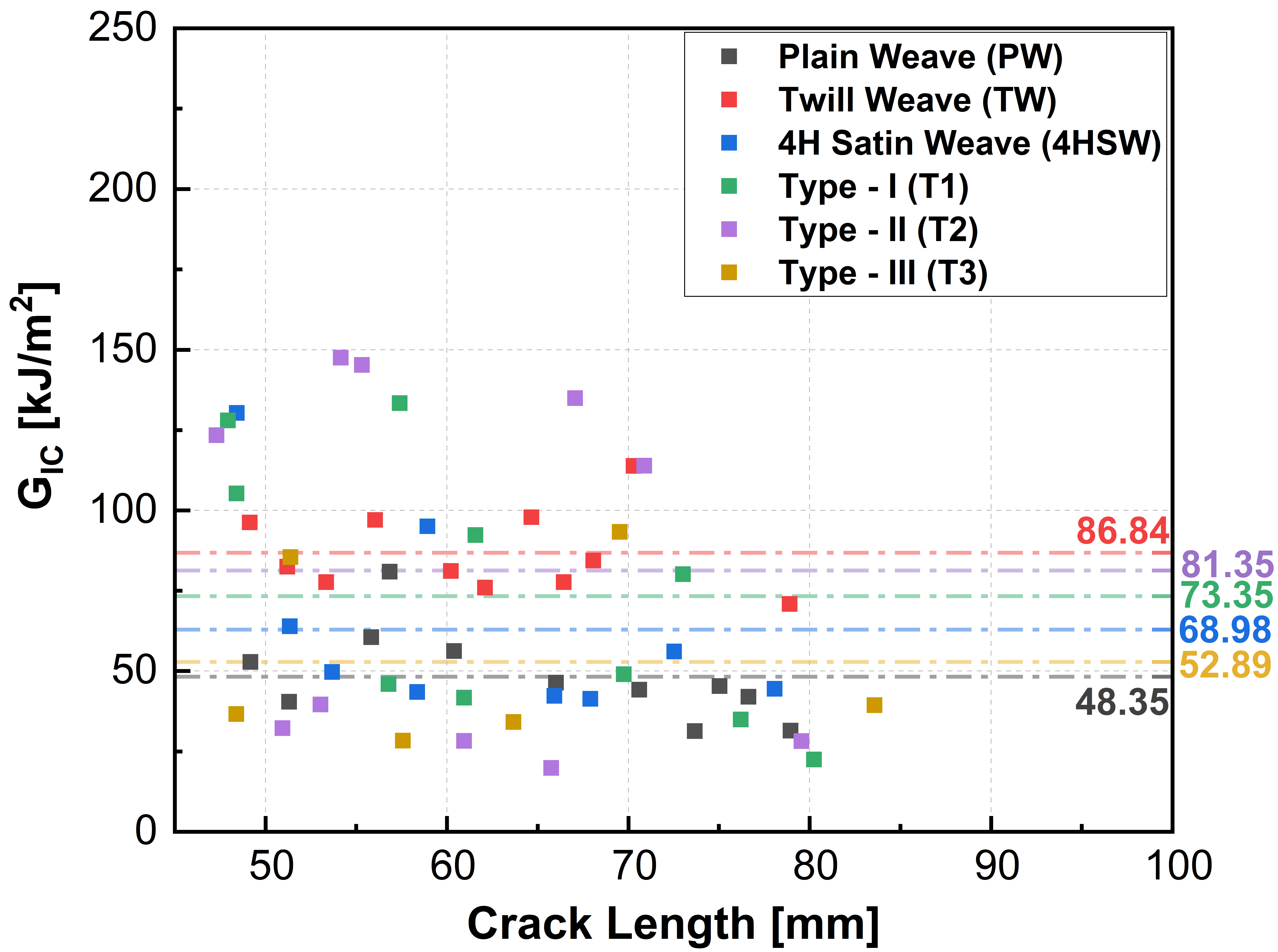}
\label{img:resfractGva}
}
\caption{Representative load-displacement responses for (a) all six weave patterns; (b) Plain weave composites; and (c) Type-II weave composites, showcasing uneven load drops in Type-II weave composites. Representative crack propagation rate vs time plot for all (d) uniform weave composites and (e) architected weave composites demonstrating tortuous crack propagation in architected weave composites. Representative R-curves for the fracture energy tests for one sample for each weave pattern (f) - uniform and architected weave composites.}
\end{figure}

Representative load-displacement responses from compact tension tests for all six weave composites are summarized in \textbf{Figure} \ref{img:resfractLvD}. The representative load-displacement responses for two cases - uniform plain weave and Type-II composites are shown in \textbf{Figures} \ref{img:resfractLvDPW} and \ref{img:resfractLvDT2}, respectively. We observe that the load drops with crack propagation were consistent and uniformly spaced for plain weave compared to the Type-II response, which manifested large sudden drops. 

\textbf{Figures} \ref{img:ratevstime1} and \ref{img:ratevstime2} show the crack propagation rates for uniform and architected weave composites, respectively. We see that the crack propagation rate values and their variation with time (acceleration) are lower compared to architected weave composites. These large drops and higher variations in crack propagation rate corresponded to areas where the weave sub-patterns transitioned, leading to higher energy release for crack propagation. It is worth noting that for Type-II and Type-III weave composites, the crack propagation in the regions with plain weave sub-patterns was similar to the uniform plain weave composites. Additionally, higher variation is observed in the transition regions as shown in \textbf{Figure} \ref{img:ratevstime2}.  

\subsubsection{R-curves}\label{res:Gvsa}
The representative R-curves (i.e., ${G_{Ic}}$ versus a) derived from all six weave composites are presented in \textbf{Figure} \ref{img:resfractGva}. We observe that uniform weave composites exhibited nearly consistent  ${G_{Ic}}$ values with crack propagation spread over a narrow value range. Within this category, twill weave composites demonstrated the highest average ${G_{Ic}}$ values, followed by satin weave and plain weave composites. This behavior arises from different fracture mechanisms observed in plain, twill, and satin weave composite, and a competition between these mechanisms. In plain weave (lowest $n_g$) composites, predominantly fiber fracture was evident, whereas satin weave (highest $n_g$) composites also demonstrated fiber pull-out. Due to inherently higher yarn mobility, fiber pull-out during fracture occurred in satin weave composites. The fiber pull-out and fiber fracture resulted in a higher average ${G_{Ic}}$ value for satin weave composites compared to plain weave counterparts. Conversely, in twill weave (intermediate $n_g$) composites, although fiber pull-out was observed, the reduced yarn mobility due to a higher number of interlaced regions (lower $n_g$) compared to satin weave, required more energy for fiber pull-out. Therefore, ${G_{Ic}}$ values for twill weave composites significantly increased compared to satin weave composites.  

For architected weave composites, the R-curve showed a counter-intuitive path. Transition regions between different sub-patterns exhibited a sudden increase in ${G_{Ic}}$ values, while ${G_{Ic}}$ values remained nearly consistent within the sub-patterns. Contrary to our expectations, regions with longer floats (with high $n_g$) displayed notably high ${G_{Ic}}$ values when transitioning from regions with shorter floats (with low $n_g$). This was observed even though the uniform satin weave (higher $n_g$) displayed lower ${G_{Ic}}$ than the twill weave (relatively lower $n_g$) due to their higher yarn mobility. However, in architected weave composites, the longer floats are still connected to the plain weave region, which restricts the yarn mobility of the floats. Therefore, for architected patterns, the energy required for fiber pull-out and fracture will be higher resulting in a higher ${G_{Ic}}$ value. Such insights suggest strategies to design weave patterns with sub-patterns to arrest crack growth and delay the final failure.    

\begin{figure}[h!]
\centering
\subfigure[]{
\includegraphics[width=0.46\textwidth]{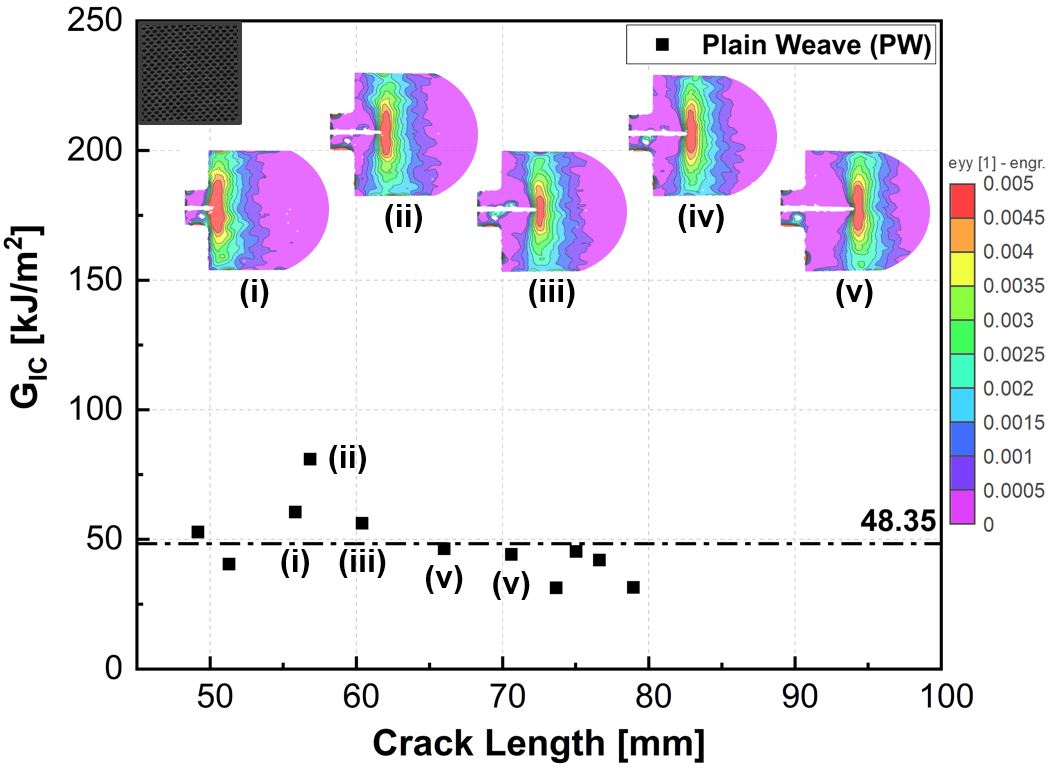}
\label{img:resGvAPW}
}
\subfigure[]{
\includegraphics[width=0.46\textwidth]{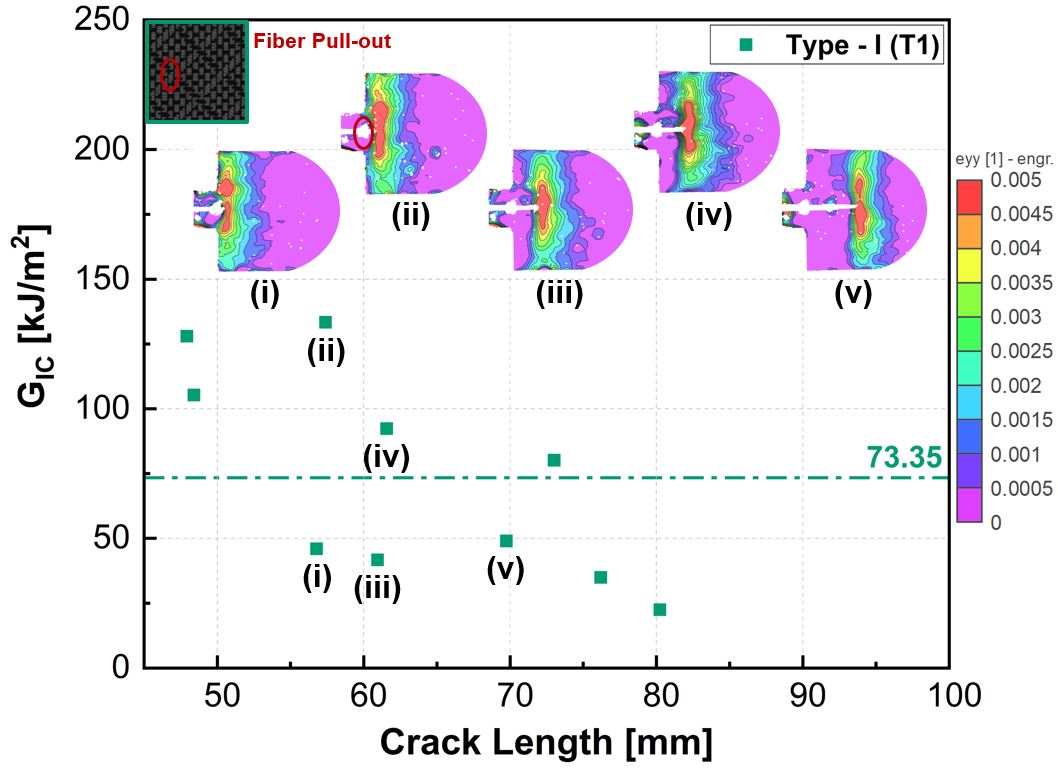}
\label{img:resGvAT1}
}
\subfigure[]{
\includegraphics[width=0.46\textwidth]{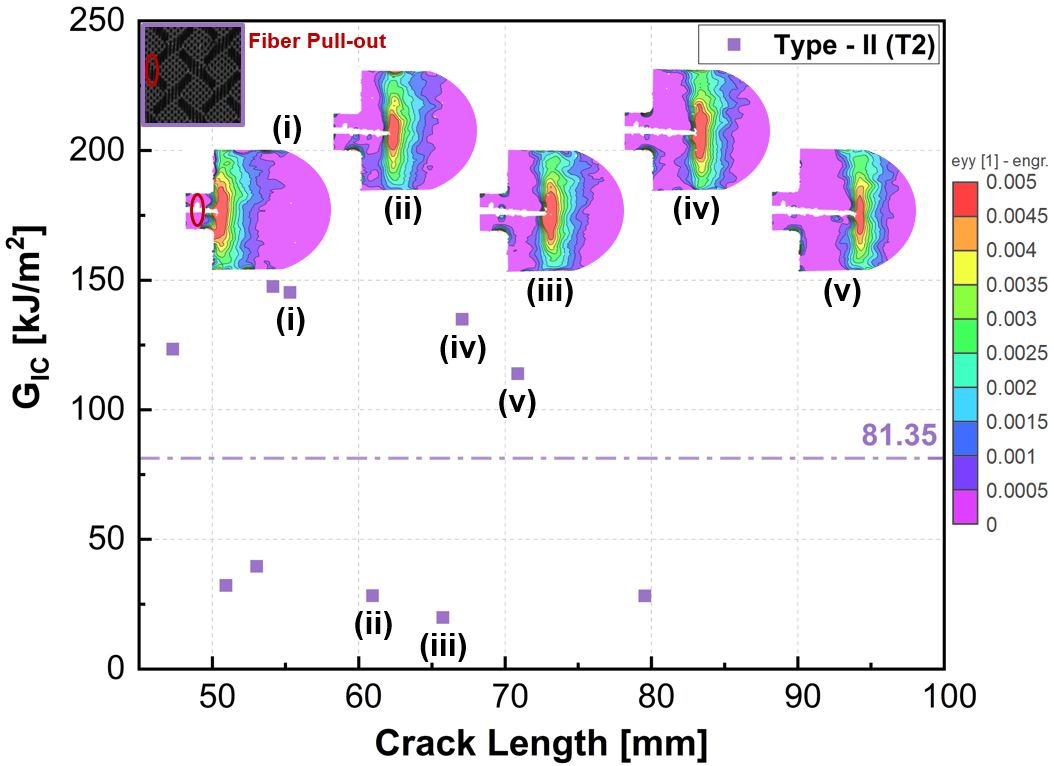}
\label{img:resGvAT2}
}
\subfigure[]{
\includegraphics[width=0.46\textwidth]{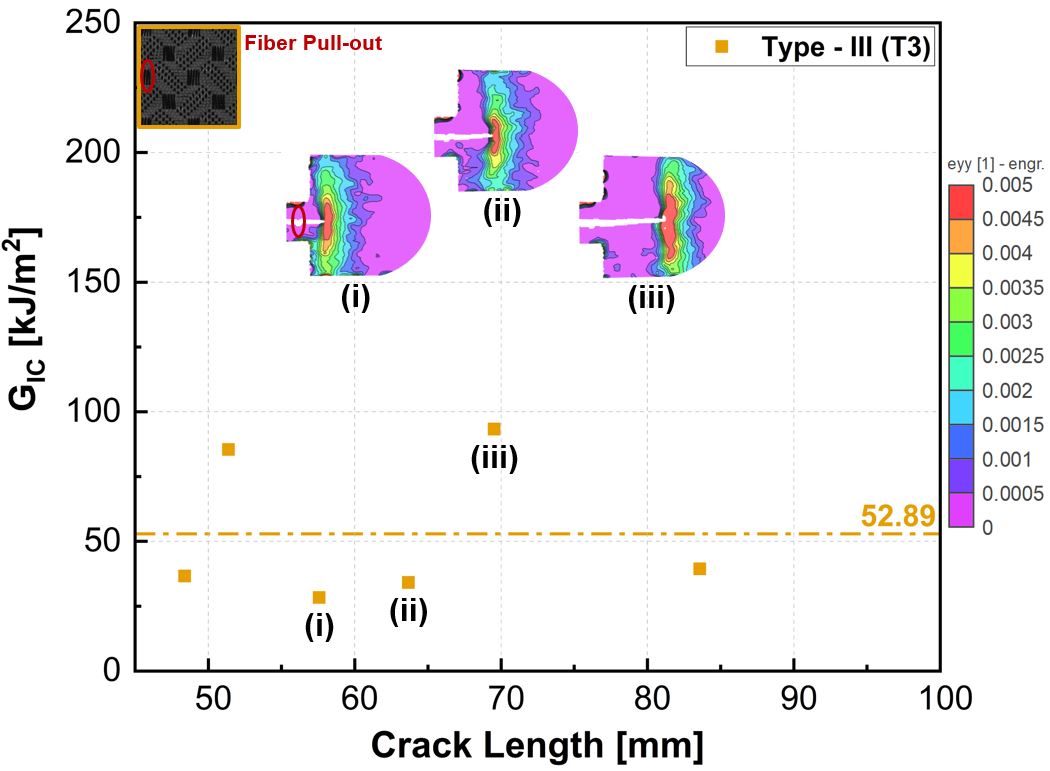}
\label{img:resGvAT3}
}
\subfigure[]{
\includegraphics[width=0.6\textwidth]{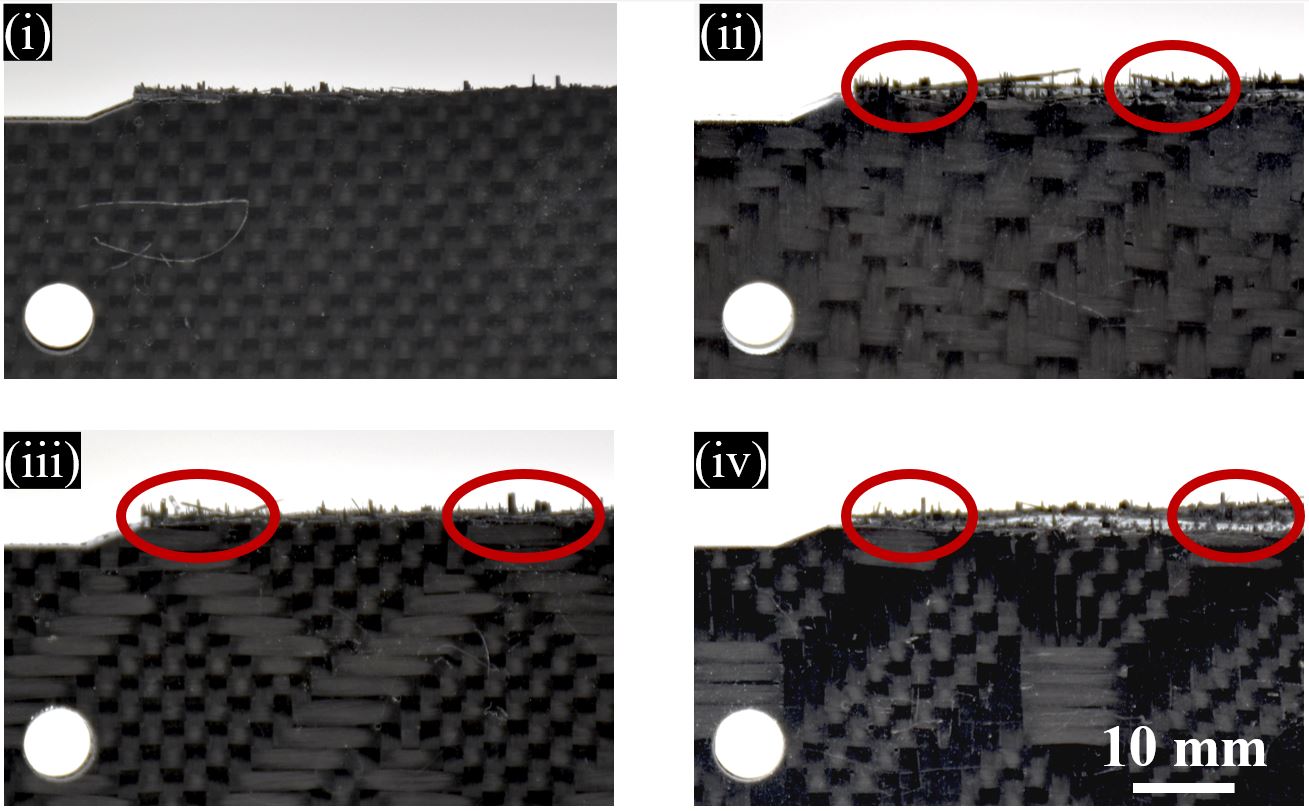}
\label{img:resmorph}
}
\caption{DIC strain maps in the loading direction (${\epsilon_{yy}}$) highlighting the fracture process zone (FPZ) in red for: (a) Plain; (b) Type-I; (c) Type-II; and (d) Type-III weave composites. Architected weave composites exhibited significant variations in FPZ at transition zones and fiber pull-out in the float regions. (e) Images of fracture surfaces of (i) Plain; (ii) Type-I; (iii) Type-II; and (iv) Type-III weave composites. Fiber pull-out (red circles) and tortuous crack paths for architected weave composites can be observed.}
\label{img:resGvADIC}
\end{figure}

We determined Digital Image Correlation (DIC) strain maps ahead of the crack tip at crack lengths between 55 mm and 70 mm for compact tensile specimens of plain weave and architected weave composites, presented in \textbf{Figure} \ref{img:resGvADIC}. From these maps, we observed two vital points to describe the differences in crack propagation between uniform and architected weave composites. For plain weave composites, the fracture process zone (FPZ), which is characterized by the strain concentration area ahead of the crack tip, showed no significant changes in size and shape. However, for architected weave composites, the FPZ exhibited considerable variations in both size and shape. The varying FPZ in these composites can be observed in \textbf{Figure} \ref{img:resGvADIC}(b-d). Although we did not see a direct correlation between FPZ size and fracture energy, this variation is an important observation to highlight the impact of multiple sub-patterns in woven composites.

We also noted fiber pull-out in all architected weave composites as marked in the red box in \textbf{Figure} \ref{img:resGvADIC}. This phenomenon occurs due to the presence of sub-patterns with higher yarn mobility, leading to fiber pull-out before fiber fracture. In contrast, in sub-patterns with lower yarn mobility, only fiber fracture was observed. These differences in failure morphologies for the entire crack propagation through these composites are also highlighted in \textbf{Figure} \ref{img:resmorph}. These varying fracture mechanisms, resulting from spatially varying sub-patterns, caused tortuous crack propagation in architected weave composites compared to uniform weave composites. These findings highlight a clear difference in fracture mechanisms between uniform and architected weave composites, demonstrating the role of weave architectures in influencing fracture energy and crack path tortuosity.

\subsubsection{Architected weave factors versus composite fracture}
Table~\ref{tab:weavepar} summarizes the transition, area, and skewness factors for the three architected weave patterns considered and their corresponding fracture energy ${G_{Ic}}$ (minimum/maximum and average) values. Fracture energy values of the uniform weave composites are further summarized in Table~\ref{tab:weavepar}. These ${G_{Ic}}$ values are obtained from one complete crack propagation in three samples for three uniform and three architected weave composites considered in this paper. 

From the above discussion, higher jumps in ${G_{Ic}}$ at the transition regions were attributed to the change in fracture mechanisms across different weave patterns. Therefore, it indicates that weave patterns and their positions as sub-patterns play a vital role in enhancing fracture energy in composites. A higher transition factor $\delta$ will cause a more evident change in fracture mechanisms, leading to a higher jump in ${G_{Ic}}$ at the transition region. Therefore, the weave sub-patterns should exhibit a higher $\delta$ to achieve woven composites with higher fracture energy. Additionally, for further improvement in ${G_{Ic}}$, we observed that a higher area factor $\rho$ and lower skewness factor $\sigma$ is favorable. This is because a larger area of lower $n_g$ will better facilitate a change in fracture mechanism. To highlight the influence of skewness, we compare Type-II and Type-III weaves. In Type-II, the centroid of the plain weave (lower $n_g$) region is far from the 5-H Satin weave due to less skewness. Conversely, in Type-III weave, this distance is smaller due to higher skewness, resulting in less drastic changes in fracture mechanisms. So, in summary, \textit{architected weave patterns with large and less skewed lower $n_g$ regions surrounded by relatively higher $n_g$ sub-patterns but smaller regions are likely to result in a higher increase in fracture energy.}

\begin{table}[h!]
\centering
\renewcommand{\arraystretch}{1.2}
\caption{Summary of fracture energy, ${G_{Ic}}$, (minimum-maximum and average) for uniform and architected weave patterns considered in this study. Factors highlighted in \textcolor{teal}{\textbf{green}} have a positive correlation with fracture energy, whereas factor in \textcolor{red}{\textbf{red}} is inversely correlated to fracture energy. These factors do not exist for uniform weaves, hence, those entries are left blank. Uniform weave composites demonstrated less significant jumps in \mbox{$G_{Ic}$} compared to architected weave composites.} 
\resizebox{15 cm}{!}{%
\begin{tabular}{lp{2cm}p{2cm}p{2cm}p{3cm}p{3cm}}
\hline
    \textbf{Weave pattern} &  \textcolor{teal}{\textbf{Transition factor ($\delta$)}} & \textcolor{teal}{\textbf{Area factor ($\rho$)}}  & \textcolor{red}{\textbf{Skewness factor ($\sigma$)}} & \textbf{Minimum/ Maximum ${G_{Ic} (KJ/m^2)}$} & \textbf{Average $\pm \hspace{0.1cm} range$   ${G_{Ic} (KJ/m^2)}$ } \\ \hline
    \textit{Type - I} & 7.23 & 0.96 & 1 & 22.52 / 149.277 & 74.87 $\pm$ 34.85  \\ 
    \textit{Type - II} & 10.5 & 0.54 & 1 & 19.90 / 157.957 & 69.23 $\pm$ 40.79   \\ 
    \textit{Type - III} & 10.5 & 0.24 & 2.67 & 22.33 / 137.155 & 55.78 $\pm$ 30.97   \\ 
    \textit{Plain weave} & -- & -- & -- & 23.36 / 72.546 & 43.83 $\pm$ 10.94  \\
    \textit{Twill weave} & -- & -- & -- & 34.18/ 113.915 & 75.48 $\pm$ 19.53  \\
    \textit{4-H Satin weave} & -- & -- & -- & 41.378 / 125.55 & 62.45 $\pm$ 27.44  \\\hline
\end{tabular}%
}
\label{tab:weavepar}
\end{table}

\section{Conclusions}\label{conc}
This study characterizes tensile properties and intralaminar Mode-I fracture energy of woven composites, emphasizing the influence of weave architectures. We show that different weave architectures can be strategically combined to tailor the mechanical performance of woven composites. We also elucidate how these weave architectures impact crack propagation in woven composites and offer avenues to designing architectures with higher fracture energy. To that end, we examined three uniform and three architected weave composites, and observed distinct behaviors. Uniform weave composites demonstrated nearly consistent fracture energy values throughout crack propagation. On the other hand, the architected weave composites exhibited higher fracture energy values at the transition regions between sub-patterns. 

\paragraph{Tensile}
We showed that for {\bf uniform weave} patterns, satin weave composites had the highest tensile properties (modulus and strength) compared to twill and plain weave composites due to the presence of fewer regions of undulations in satin weave. Therefore, a weave pattern can be designed with more float regions by increasing the ${n_g}$ value to improve the mechanical properties. However, it is worth noting that the increase in tensile properties from twill to 4-H satin weave is marginal, indicating that the mechanical properties will likely plateau with a further increase in ${n_g}$ value. In {\bf architected weave} composites, we observed that the weakest sub-pattern (lowest $n_g$) controlled the tensile properties. For example, Type-II and Type-III composites showed properties similar to those of plain weave composites due to the presence of plain weave in them. These insights can be used by strategically choosing sub-patterns to create architected weave patterns with higher tensile properties. While the tensile properties of architected weave composites ranged between plain weave and satin weave composites, they exhibited high fracture energy (${G_{Ic}}$) jumps at the transition regions between different weave sub-patterns.

\paragraph{Fracture}
In terms of the fracture energy, ${G_{Ic}}$, the {\bf uniform weave} composites exhibited consistent load drops and crack propagation rates, resulting in nearly consistent fracture energy values with crack propagation. We also showed that different uniform weave architectures, like plain, twill, satin, etc. influenced the failure mechanisms, hence affecting the ${G_{Ic}}$ values. Plain weave exhibited predominantly fiber breakage, whereas satin weave composites had fiber pull-out in addition to fiber breakage. However, due to higher yarn mobility in satin weave, the increase in ${G_{Ic}}$ was not large compared to plain weave composites. On the other hand, the twill weave composites also exhibited fiber pull-out and fiber breakage, but due to lesser yarn mobility, these composites had the highest average ${G_{Ic}}$ values. These findings show that different uniform weave architectures result in different failure mechanisms that can be tuned to suit the design requirements. 

Our investigations also unveiled counter-intuitive R-curves for {\bf architected weave} composites. We showed, for the first time, that ${G_{Ic}}$ values increased at the transition regions from one sub-pattern to another. Also, as the float regions with higher $n_g$ were connected to weave patterns with lower ${n_g}$ value, the yarn mobility decreased. This also played a vital role in increasing the ${G_{Ic}}$ value near the float regions in architected weave composites.

We also proposed three geometrical factors for {\bf architected weaves} to characterize their effects on the fracture energy of woven composites. Higher transition ($\delta$) and area ($\rho$) factors and lower skewness factor ($\sigma$) are favorable to promote change in failure mechanisms when a crack propagates from one sub-pattern to another, resulting in higher fracture energy. However, higher $\delta$ would result in localized areas of stress concentrations under tensile loading. 

Our findings not only deepen our fundamental understanding of the relationship between crack propagation and weave architectures but also pave the way to design novel weave patterns with tunable mechanical performance and controlled damage mechanisms. These woven fabrics can be reinforced to create safer and longer-lasting woven composite structures with higher damage tolerance and improved mechanical properties for diverse applications in the aerospace, automotive, or marine industries.

\section*{Funding}
This research was partly funded by the NSF CAREER award through {\em{Mechanics of Materials and Structures (MOMS) Program}} [\#: 2046476].


\section*{Author Contributions}
Author H.R.T. contributed to the conceptualization of the methodology, formal analysis, investigating, visualization, verification, and writing -- preparing the original draft. Authors J.C. and K.P. contributed to the manufacturing and investigating. Author P.P. contributed to the conceptualization of the methodology, writing -- reviewing and editing, visualization, verification, supervision, project administration, and funding acquisition.

\section*{Data Availability}
The data that support the findings of this study are available from the corresponding author, PP, upon reasonable request.


{\footnotesize
\bibliographystyle{unsrt}
\bibliography{Woven_bib}

\begin{thebibliography}{10}

\bibitem{onal2007modeling}
L.~Onal and S.~Adanur.
\newblock Modeling of elastic, thermal, and strength/failure analysis of
  two-dimensional woven composites — a review.
\newblock 2007.

\bibitem{lisle2015damage}
T.~Lisle, C.~Bouvet, M.-L. Pastor, T.~Rouault, and P.~Marguer{\`e}s.
\newblock Damage of woven composite under tensile and shear stress using
  infrared thermography and micrographic cuts.
\newblock {\em Journal of materials science}, 50:6154--6170, 2015.

\bibitem{Bakar2013}
I.~A.~A. Bakar, O.~Kramer, S.~Bordas, and T.~Rabczuk.
\newblock Optimization of elastic properties and weaving patterns of woven
  composites.
\newblock {\em Composite Structures}, 100:575--591, 2013.

\bibitem{ullah2015characterisation}
H.~Ullah, A.~R. Harland, and V.~V. Silberschmidt.
\newblock Characterisation of mechanical behaviour and damage analysis of {2D}
  woven composites under bending.
\newblock {\em Composites Part B: Engineering}, 75:156--166, 2015.

\bibitem{erol2017effects}
O.~Erol, B.~M. Powers, and M.~Keefe.
\newblock Effects of weave architecture and mesoscale material properties on
  the macroscale mechanical response of advanced woven fabrics.
\newblock {\em Composites Part A: Applied Science and Manufacturing},
  101:554--566, 2017.

\bibitem{zhou2018multi}
L.~C. Zhou, M.~Chen, C.~Liu, and H.~A. Wu.
\newblock A multi-scale stochastic fracture model for characterizing the
  tensile behavior of {2D} woven composites.
\newblock {\em Composite Structures}, 204:536--547, 2018.

\bibitem{vaidya2023performance}
U.~Vaidya, B.~Thattaiparthasarthy, M.~Janney, M.~Mauhar, K.~Graham, and
  E.~Cates.
\newblock Performance of hybrid innegra-carbon fiber composites.
\newblock {\em Scientific Reports}, 13(1):20876, 2023.

\bibitem{ishikawa1982}
T.~Ishikawa and T.~W. Chou.
\newblock Stiffness and strength behaviour of woven fabric composites.
\newblock {\em Journal of Materials Science}, 17:3211--3220, 1982.

\bibitem{ishikawa1982elastic}
T.~Ishikawa and T.~W. Chou.
\newblock Elastic behavior of woven hybrid composites.
\newblock {\em Journal of composite materials}, 16(1):2--19, 1982.

\bibitem{ishikawa1983one}
T.~Ishikawa and T.~W. Chou.
\newblock One-dimensional micromechanical analysis of woven fabric composites.
\newblock {\em AIAA journal}, 21(12):1714--1721, 1983.

\bibitem{naik1995analytical}
N.~K. Naik and V.~K. Ganesh.
\newblock An analytical method for plain weave fabric composites.
\newblock {\em Composites}, 26(4):281--289, 1995.

\bibitem{scida1998prediction}
D.~Scida, Z.~Aboura, M.~L. Benzeggagh, and E.~Bocherens.
\newblock Prediction of the elastic behaviour of hybrid and non-hybrid woven
  composites.
\newblock {\em Composites Science and Technology}, 57(12):1727--1740, 1998.

\bibitem{qian2023numerical}
Y.~Qian, Q.~Bao, Z.~Li, T.~Xia, Z.~Yang, and Z.~Lu.
\newblock Numerical investigation on the mechanical behaviors of {2D} woven
  composites under complex in-plane stress states.
\newblock {\em Composite Structures}, 315:117008, 2023.

\bibitem{Feng2022}
H.~Feng, S.~P. Subramaniyan, H.~Tewani, and P.~Prabhakar.
\newblock Physics-constrained neural network for design and feature-based
  optimization of weave architectures.
\newblock 2023.

\bibitem{Xu2017}
W.~Xu and A.~M. Waas.
\newblock Fracture toughness of woven textile composites.
\newblock {\em Engineering Fracture Mechanics}, 169:184--188, 2017.

\bibitem{murdani2009fatigue}
A.~Murdani, C.~Makabe, and M.~Fujikawa.
\newblock Fatigue and fracture behavior in notched specimens of {C/C} composite
  with fine-woven carbon fiber laminates.
\newblock {\em Carbon}, 47(14):3355--3364, 2009.

\bibitem{Katafiasz2019}
T.~J. Katafiasz, L.~Iannucci, and E.~S. Greenhalgh.
\newblock Development of a novel compact tension specimen to mitigate premature
  compression and buckling failure modes within fibre hybrid epoxy composites.
\newblock {\em Composite Structures}, 207:93--107, 2019.

\bibitem{Dalli2019}
D.~Dalli, G.~Catalanotti, L.~F. Varandas, B.~G. Falzon, and S.~Foster.
\newblock {Mode I} intralaminar fracture toughness of {2D} woven carbon fibre
  reinforced composites: A comparison of stable and unstable crack propagation
  techniques.
\newblock {\em Engineering Fracture Mechanics}, 214:427--448, 2019.

\bibitem{cheng2022study}
P.~Cheng, Y.~Peng, K.~Wang, Y.~Q. Wang, and C.~Chen.
\newblock Study on intralaminar crack propagation mechanisms in single-and
  multi-layer {2D} woven composite laminate.
\newblock {\em Mechanics of Advanced Materials and Structures},
  29(25):4310--4318, 2022.

\bibitem{donadon2007intralaminar}
M.~V. Donadon, B.~G. Falzon, L.~Iannucci, and J.~M. Hodgkinson.
\newblock Intralaminar toughness characterisation of unbalanced hybrid plain
  weave laminates.
\newblock {\em Composites Part A: Applied Science and Manufacturing},
  38(6):1597--1611, 2007.

\bibitem{rokbi2011}
M.~Rokbi, H.~Osmani, N.~Benseddiq, and A.~Imad.
\newblock On experimental investigation of failure process of woven-fabric
  composites.
\newblock {\em Composites Science and Technology}, 71(11):1375--1384, 2011.

\bibitem{boyina2014mixed}
Dhatreyi Boyina, Anuradha Banerjee, and R~Velmurugan.
\newblock Mixed-mode translaminar fracture of plain-weave composites.
\newblock {\em Composites Part B: Engineering}, 60:21--28, 2014.

\bibitem{liu2008fracture}
Q.~Liu and M.~Hughes.
\newblock The fracture behaviour and toughness of woven flax fibre reinforced
  epoxy composites.
\newblock {\em Composites Part A: Applied Science and Manufacturing},
  39(10):1644--1652, 2008.

\bibitem{Blanco2014}
N.~Blanco, D.~Trias, S.~T. Pinho, and P.~Robinson.
\newblock Intralaminar fracture toughness characterisation of woven composite
  laminates. {P}art {II}: Experimental characterisation.
\newblock {\em Engineering Fracture Mechanics}, 131:361--370, 2014.

\bibitem{Plain}
FibreGlast~Development Corp.
\newblock {3K, Plain Weave Carbon Fiber}.

\bibitem{Twill}
FibreGlast~Development Corp.
\newblock {3K, 2 x 2 Twill Weave Carbon Fiber}.

\bibitem{MBoss}
FibreGlast~Development Corp.
\newblock {3K, Textral Weave (M-Boss Pattern) Carbon Fiber}.

\bibitem{Diam}
FibreGlast~Development Corp.
\newblock {3K, Textral Weave (Diamondplate Pattern) Carbon Fiber}.

\bibitem{Rosw}
FibreGlast~Development Corp.
\newblock {3K, Textral Weave (Roswell Pattern) Carbon Fiber}.

\bibitem{osada2003initial}
T.~Osada, A.~Nakai, and H.~Hamada.
\newblock Initial fracture behavior of satin woven fabric composites.
\newblock {\em Composite structures}, 61(4):333--339, 2003.

\bibitem{Tapie2016}
E.~Tapie, Y.~B. Guo, and V.~P.W. Shim.
\newblock Yarn mobility in woven fabrics - a computational and experimental
  study.
\newblock {\em International Journal of Solids and Structures}, 80:212--226,
  2016.

\bibitem{haralick1973textural}
R.~M. Haralick, K.~Shanmugam, and I.~H. Dinstein.
\newblock Textural features for image classification.
\newblock {\em IEEE Transactions on Systems, Man, and Cybernetics},
  (6):610--621, 1973.

\bibitem{zhou2021experimental}
G.~Zhou, Q.~Sun, Z.~Meng, D.~Li, Y.~Peng, D.~Zeng, and X.~Su.
\newblock Experimental investigation on the effects of fabric architectures on
  mechanical and damage behaviors of carbon/epoxy woven composites.
\newblock {\em Composite structures}, 257:113366, 2021.

\bibitem{ASTM3039}
ASTM D3039/D3039M.
\newblock {\em ASTM Standards - Standard Test Method for Tensile Properties of
  Polymer Matrix Composite Materials}, 2008.

\bibitem{ASTM399}
ASTM E399.
\newblock {\em ASTM Standards - Standard Test Method for Linear-Elastic
  Plane-Strain Fracture Toughness of Metallic Materials}, 2018.

\bibitem{Laffan2013}
M.~J. Laffan, S.~T. Pinho, and P.~Robinson.
\newblock Mixed-mode translaminar fracture of {CFRP}: Failure analysis and
  fractography.
\newblock {\em Composite Structures}, 95:135--141, 2013.

\bibitem{LAFFAN2010606}
M.~J. Laffan, S.~T. Pinho, P.~Robinson, and L.~Iannucci.
\newblock Measurement of the in situ ply fracture toughness associated with
  {Mode I} fibre tensile failure in {FRP}. {P}art {I}: data reduction.
\newblock {\em Composites Science and Technology}, 70(4):606--613, 2010.

\end{thebibliography}
}

\newpage



\end{document}